\documentclass{aa}
\usepackage{graphicx}
\usepackage{txfonts}
\usepackage[colorlinks=true, linkcolor=blue, citecolor=blue, urlcolor=blue, breaklinks=true]{hyperref}
\def\oiii{[O~{\sc iii}]/H$\beta$}
\def\nii{[N~{\sc ii}]/H$\alpha$}
\def\sii{[S~{\sc ii}]/H$\alpha$}
\def\oi{[O~{\sc i}]/H$\alpha$}
\usepackage{listings}
\usepackage{xcolor}
\usepackage{booktabs}

\begin{document} 

\title{An improved Seyfert-LINER classification line in the [N~{\sc ii}] BPT diagram}

\author{PeiZhen Cheng \inst{1}, 
XingQian Chen \inst{1}, GuiLin Liao \inst{1}, Ying Gu \inst{1}, Qi Zheng \inst{1}
\and XueGuang Zhang \inst{1}\fnmsep\thanks{E-mail: \href{mailto:xgzhang@gxu.edu.cn}{xgzhang@gxu.edu.cn}}}

\institute{School of Physical Science and Technology, Guangxi University, No. 100, Daxue East Road, Nanning, 530004, P. R. China}

\titlerunning{S-L line in the \texttt{[N~{\sc ii}]} BPT diagram}
\authorrunning{Cheng et al.}


\abstract
  {In this manuscript, an improved Seyfert-LINER classification line (= S-L line) is proposed in the [N~{\sc ii}] BPT diagram, 
based on a sample of 47,968 low redshift narrow emission line galaxies from SDSS DR16, motivated by different S-L lines reported in 
the [N~{\sc ii}] BPT diagram through different methods. The method proposed by Kewley et al. in 2006 is firstly applied, however, 
the method cannot lead to an accepted S-L line in the [N~{\sc ii}] BPT diagram. Meanwhile, the S-L lines proposed by Schawinski et 
al. in 2007 and Cid Fernandes et al. in 2010 in the [N~{\sc ii}] BPT diagram are different from each other.}
  {Therefore, it is meaningful to check which proposed S-L line is better or to determine an improved one in the [N~{\sc ii}] BPT 
diagram by a new method.}
  {In this manuscript, Seyferts and LINERs that have already been classified in the [S~{\sc ii}] and/or [O~{\sc i}] BPT diagrams 
can be visualized in the [N~{\sc ii}] BPT diagram, leading the intersection boundary of the two contour maps to be considered as the 
S-L line in the [N~{\sc ii}] BPT diagram.}
  {Rather than the previously proposed S-L lines, the new S-L line can lead to more efficient and harmonious classifications 
of Seyferts and LINERs, especially in the composite galaxy region, in the [N~{\sc ii}] BPT diagram. Furthermore, based on the 
discussed S-L lines, the number ratio of Type-2 Seyferts to Type-2 LINERs differs significantly from that of Type-1 Seyferts 
to Type-1 LINERs in the [N~{\sc ii}] BPT diagram, suggesting that about 90$\%$ of Type-2 LINERs are non-AGN-related objects, 
true Type-2 AGNs, or objects exhibiting both Seyfert and LINER characteristics.}
   {}

   \keywords{Active galactic nuclei -- Emission line galaxies -- Seyfert galaxies --  LINER galaxies}

   \maketitle
%

\section{Introduction} \label{introduction}

	The well-known BPT (Baldwin, Phillips \& Terlevich) diagrams have been widely accepted for classifying narrow emission 
line galaxies, since \citet{1981PASP...93....5B} and \citet{1987ApJS...63..295V} have clearly shown that different types of 
extragalactic objects have different lying regions in space of flux ratios of different narrow emission lines, especially flux ratios 
of [O~{\sc iii}]$\lambda$5007\AA\ to H$\beta$ versus [N~{\sc ii}]$\lambda$6584\AA\ to H$\alpha$ (the [N~{\sc ii}] BPT diagram), 
of [O~{\sc iii}]$\lambda$5007\AA\ to H$\beta$ versus [S~{\sc ii}]$\lambda$6717,6731\AA\ to H$\alpha$ (the [S~{\sc ii}] BPT diagram) 
and of [O~{\sc iii}]$\lambda$5007\AA\ to H$\beta$ versus [O~{\sc i}]$\lambda$6300\AA\ to H$\alpha$ (the [O~{\sc i}] BPT diagram).
In the 1980s, due to the small sample sizes of narrow emission line galaxies applied in the BPT diagrams, only about 140 narrow 
emission line galaxies discussed in \citet{1981PASP...93....5B} and only about 240 narrow emission line galaxies discussed 
in \citet{1987ApJS...63..295V}, the classifications for different types of narrow emission line galaxies, such as the Type-2 Seyferts 
and {H\kern-0.1em\textsc{ii}} galaxies, are clear enough. However, with the very rapid growth in the number of narrow emission line 
galaxies, there are no apparent classification lines between different types of galaxies in the BPT diagrams. Therefore, how to give 
efficient classifications (or classification lines between different types of galaxies) in the BPT diagrams is a meaningful research 
subject.

	\citet{2001ApJ...556..121K} have reported the clear theoretical classification lines (hereafter as the Ke01 lines) between 
{H\kern-0.1em\textsc{ii}} regions and AGNs (Active Galactic Nuclei) in the three BPT diagrams, after considering narrow emission 
line ratios theoretically determined by ionization photons generated by the PEGASE v2.0 and STARBURST99 codes. \citet{2003MNRAS.346.1055K} 
have proposed an empirical classification line (hereafter as the Ka03 line) in the [N~{\sc ii}] BPT diagram through a large sample of 
22,623 AGNs from the SDSS (Sloan Digital Sky Survey), leading to an efficient classification of composite galaxies. 
\cite{2006MNRAS.371..972S} have proposed a classification line, lower than the Ka03 line, between pure normal star-forming galaxies 
and AGNs in the [N~{\sc ii}] BPT diagram, based on a sample of 20,000 galaxies from the SDSS, considering photoionization models 
informed by synthetic stellar radiation and a broken power-law spectrum for AGNs. \citet{2006MNRAS.372..961K} have proposed the 
widely accepted classification lines between Seyferts (commonly, the Type-2 Seyferts) and LINERs (Low Ionization Nuclear Emission-line 
Regions) in the [S~{\sc ii}] and [O~{\sc i}] BPT diagrams based on a sample of 85,224 emission-line galaxies from the SDSS, but not 
in the [N~{\sc ii}] BPT diagram. \citet{2017MNRAS.472.2808D} have applied the Gaussian Mixture Models to discuss the classification 
lines between different types of narrow emission line galaxies in the [N~{\sc ii}] BPT diagram, effectively distinguishing AGNs, 
composite galaxies and {H\kern-0.1em\textsc{ii}} regions, while struggling with Seyferts and LINERs due to overlap. 
\citet{2018MNRAS.478.3177T} have proposed that AGNs can be well distinguished from star-forming galaxies through a machine learning 
approach. \citet{2020ApJ...905...97Z} have verified the classifications of AGNs and {H\kern-0.1em\textsc{ii}} galaxies in the BPT 
diagrams based on the t-SNE technique applied to 35,857 local narrow emission-line galaxies. \cite{2021ApJ...922..156A,2023MNRAS.526.4455A} 
have improved the distinctions between AGNs and star-forming galaxies by truncating the lower portion of the Ka03 line with a straight 
line at log(\nii) = $-$0.35. A comprehensive review of emission line-based classification methods can be found in \citet{2019ARA&A..57..511K}.

	At the current stage, there are clear classification lines between {H\kern-0.1em\textsc{ii}} regions and AGNs in all the three 
BPT diagrams. However, Seyferts and LINERs are well classified only in the [S~{\sc ii}] and [O~{\sc i}] BPT diagrams, since these two 
diagrams show two distinct branches above the Ke01 line. In contrast, the [N~{\sc ii}] BPT diagram shows only one thick branch at the 
upper right side of the Ke01 line, due to the [N~{\sc ii}]/H$\alpha$ ratio not offering enough distinctions between these two branches 
\citep{2006MNRAS.371..972S,2019ARA&A..57..511K}. This distinction, as shown in our BPT diagrams in the following sections, leads to 
only Seyferts and LINERs with apparent [S~{\sc ii}] and/or [O~{\sc i}] narrow forbidden emission lines being well classified. Considering 
the number of galaxies with [N~{\sc ii}] emission lines far exceeds that of those with [S~{\sc ii}] or [O~{\sc i}] narrow emission lines, 
classifying Seyferts and LINERs in the [N~{\sc ii}] BPT diagram will undoubtedly provide richer samples by including more galaxies with 
weak [S~{\sc ii}] or [O~{\sc i}] but apparent [N~{\sc ii}] emission lines, and facilitate the smooth exclusion of LINERs from Seyferts 
in future studies to test the unified model of AGNs. Therefore, it is a meaningful subject to determine a classification line in the 
[N~{\sc ii}] BPT diagram to effectively classify Seyferts and LINERs, due to part of LINERs probably not AGN-related objects.

	The concept of LINERs is firstly defined by \cite{1980A&A....87..152H}, and \citet{1996ApJ...462..183H} improve their definitions. 
However, there is still controversy about the physical mechanism of LINERs. To begin with, some LINERs are powered by black hole accretion 
systems, indicating clear AGN-related activity. \citet{2019ApJ...876...12A} have concluded that the proportion of LINERs with X-ray 
detection is similar to that of AGNs (Seyferts). Moreover, \citet{2023ApJ...943..174A} have found that LINERs detected in X-rays are 
not significantly different from those undetected, supporting the idea that many LINERs are indeed genuine AGNs to some extent. In 
contrast, many researchers tend to believe that the dominant mechanism in some LINERs is non-AGN-related. \cite{2009ASPC..408..122C} 
have proposed that most LINER-like systems in the SDSS should not be included in any census of actively accreting black holes as 
they are consistent with being retired galaxies where ionizing photons are produced by aging stars. \cite{2013A&A...558A..43S} have 
found that the radial emission-line surface brightness distribution of 48 galaxies with LINER-like emission, based on integral 
field spectroscopy data from the Calar Alto Legacy Integral Field Area survey, is inconsistent with ionization by a central point 
source (AGN illumination), suggesting that LINERs cannot be solely attributed to AGNs. Furthermore, shock wave heating models have 
been proposed to explain the differences between Seyferts and LINERs \citep{1995ApJ...455..468D,1996ASPC..103...44D, 
2011ApJ...734...87R,2014ApJ...781L..12R}. At least part of the LINERs should not be considered as genuine AGNs. Moreover, it is 
likely that multiple physical processes contribute to LINER activity, making it unsuitable to describe LINERs with a single 
mechanism \citep{2001ApJS..132..211C,2012ApJ...747...61Y,2013A&A...558A..34B,2018MNRAS.476.2457C}. Further discussions on different 
physical mechanisms of LINERs can be found in \cite{2008ARA&A..46..475H,2014ARA&A..52..589H,2017FrASS...4...34M}. 
\citet{1997ApJS..112..315H}, following the traditional way of naming Seyferts, classified LINERs into Type-1 and Type-2 
based on the presence or absence of broad (H$\alpha$) emission lines. As discussed in \citet{2008ARA&A..46..475H}, this observational 
distinction reflects intrinsic structural differences in the central engine, rather than orientation effects, between these two types 
of LINERs. Type-1 LINERs are associated with relatively higher accretion rates that can sustain a weak broad-line region. In contrast, 
Type-2 LINERs accrete at extremely sub-Eddington rates, under which both the broad-line region and the torus intrinsically disappear, 
rather than being obscured along the line of sight. As a result, part of LINERs do not follow the unified model of AGNs.
Whether LINERs are AGN-related or non-AGN-related is still controversial. Due to the existence of non-AGN-related LINERs, there 
should be potential effects on physical property comparisons between Type-1 AGNs and Type-2 AGNs to test the commonly applied 
unified model of AGNs, as proposed and discussed in \citet{1993ARA&A..31..473A,1995PASP..107..803U,2015ARA&A..53..365N}, if 
non-AGN-related LINERs are included in Type-2 AGNs.

	   Great progress has been made in classifying Seyferts and LINERs within the BPT diagrams. However, unlike the unanimous 
S-L lines in the [S~{\sc ii}] and [O~{\sc i}] BPT diagrams reported in \citet{2006MNRAS.372..961K}, different S-L lines have been 
presented in the [N~{\sc ii}] BPT diagram in the literature. \citet{1997ApJS..112..315H} have proposed a horizontal S-L line in the 
[N~{\sc ii}] BPT diagram, characterizing LINERs having \oiii $<$ 3 and \nii $>$ 0.6. However, as pointed out in 
\cite{2008ARA&A..46..475H}, the horizontal S-L line in \citet{1997ApJS..112..315H} has no strict physical significance. Therefore, 
there are no further considerations on the S-L line in \citet{1997ApJS..112..315H} in this manuscript. \citet{2007MNRAS.382.1415S} 
have proposed a diagonal S-L line (hereafter as the Sc07 line) in the [N~{\sc ii}] BPT diagram. \cite{2010MNRAS.403.1036C} have 
proposed one other diagonal S-L line (hereafter as the Fe10 line) in the [N~{\sc ii}] BPT diagram, which is different from the 
Sc07 line. Due to the lack of a harmonious S-L line in the [N~{\sc ii}] BPT diagram, to check whether a definitive S-L line can 
be determined in the [N~{\sc ii}] BPT diagram is the main objective of the manuscript.

	In this manuscript, Section \ref{sql} presents the data samples. Section \ref{sec:main} elucidates the main results on determining 
the improved classification line between Seyferts and LINERs in the [N~{\sc ii}] BPT diagram. Section \ref{sec:discussions} presents 
the necessary discussions. Section \ref{sec:Conclusion} provides the main summary and conclusions. In this manuscript, the cosmological 
parameters of $H_{0}~=~70{\rm km\cdot s^{-1}}{\rm Mpc}^{-1}$, $\Omega_{\Lambda}~=~0.7$ and $\Omega_{m}~=~0.3$ have been adopted.

\section{Data samples} \label{sql}

\begin{figure}
	\centering
	\includegraphics[width=\columnwidth]{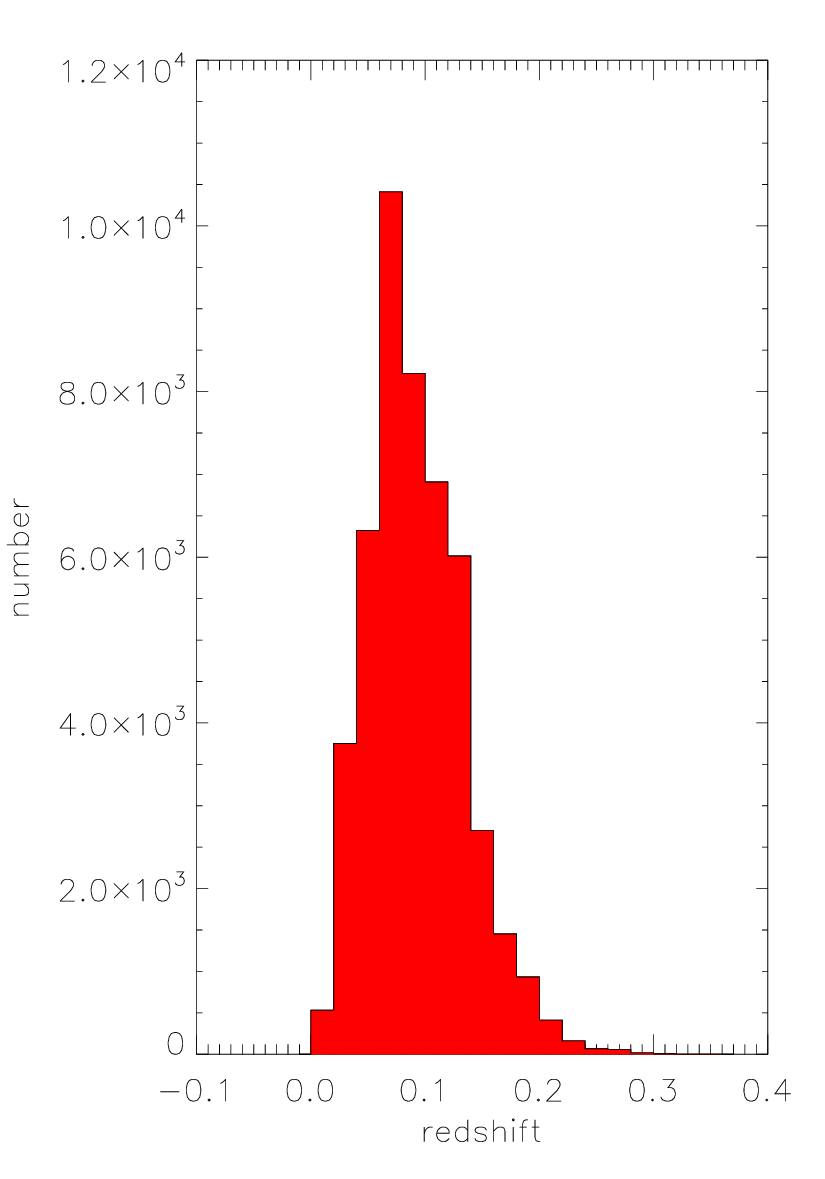}
	\caption{This figure demonstrates the redshift distribution of 47,968 low redshift narrow emission line galaxies with 
reliable narrow emission lines from SDSS DR16 using the SQL Search query. The redshifts range from 0.0002 to 0.3469.}
	\label{fig:Z}
\end{figure}

	   In order to determine the S-L line in the [N~{\sc ii}] BPT diagram, narrow emission line galaxies are firstly collected. 
Similar to what we have recently done in \citet{2022ApJS..260...31Z,2023ApJS..267...36Z,2024ApJ...964..141Z}, all the narrow emission 
line galaxies are selected using the SQL Search query\footnote{\href{https://skyserver.sdss.org/dr16/en/tools/search/sql.aspx}{https://skyserver.sdss.org/dr16/en/tools/search/sql.\allowbreak aspx}} from the SDSS 
DR16 (Data Release 16; \citealt{2020ApJS..249....3A}). Detailed conditions and query can be found in 
Appendix \ref{SQL conditions}. Then, the SQL Search query results in 47,968 low redshift narrow emission line galaxies (Sample 1) 
with reliable narrow emission lines from SDSS DR16, with redshifts ranging from 0.0002 to 0.3469 as shown in Fig.~\ref{fig:Z}.

	It is worth mentioning that, through the similar SQL Search queries, considering the Ke01 lines described by the flux ratios of 
different narrow emission lines in the BPT diagrams, 14,321 galaxies lying above the Ke01 line in the [N~{\sc ii}] BPT diagram can be 
collected with reliable [N~{\sc ii}] emission lines. Meanwhile, through the similar SQL Search queries, 12,250 galaxies lying above the 
Ke01 line in the [S~{\sc ii}] BPT diagram can be collected with reliable [S~{\sc ii}] emission lines, and 7,309 galaxies lying above 
the Ke01 line in the [O~{\sc ii}] BPT diagram can be collected with reliable [O~{\sc i}] emission lines. Therefore, in the BPT diagrams 
for the Seyferts and LINERs, the number of galaxies with reliable [N~{\sc ii}] emission lines is 17$\%$ (96$\%$) higher than those of 
galaxies with reliable [S~{\sc ii}] ([O~{\sc i}]) emission lines, respectively. In other words, to determine a well-defined S-L line in 
the [N~{\sc ii}] BPT diagram is more helpful for classifications of Seyferts and LINERs.

\section{Main results} \label{sec:main}

\subsection{Current S-L lines in the [N~{\sc ii}] BPT diagram}

	As the method shown and discussed in \citet{2006MNRAS.372..961K}, the S-L lines can be determined and shown in the [S~{\sc ii}] 
and [O~{\sc i}] BPT diagrams. Therefore, the same method is firstly applied to test its applicability in determining the S-L lines in the 
BPT diagrams based on our new sample of narrow emission line galaxies. Galaxy number count distributions with respect to the angle with 
log(\sii) and log(\oi) as the X-axis provide well double-peaked features similar to the Fig.~3 in \citet{2006MNRAS.372..961K}, leading 
to the similar S-L lines in the [S~{\sc ii}] and [O~{\sc i}] BPT diagrams as those reported in \citet{2006MNRAS.372..961K}, as the 
detailed discussions in Appendix~\ref{sec:appendix}. Unfortunately, the same method cannot lead to an accepted S-L line in 
the [N~{\sc ii}] BPT diagram, mainly because the galaxy number count distributions with respect to the angle with the log(\nii) as the 
X-axis are relatively even, failing to distinctly separate Seyferts from LINERs. More detailed discussions can be found in 
Appendix~\ref{sec:appendix}.

\begin{figure}
\includegraphics[width=\columnwidth]{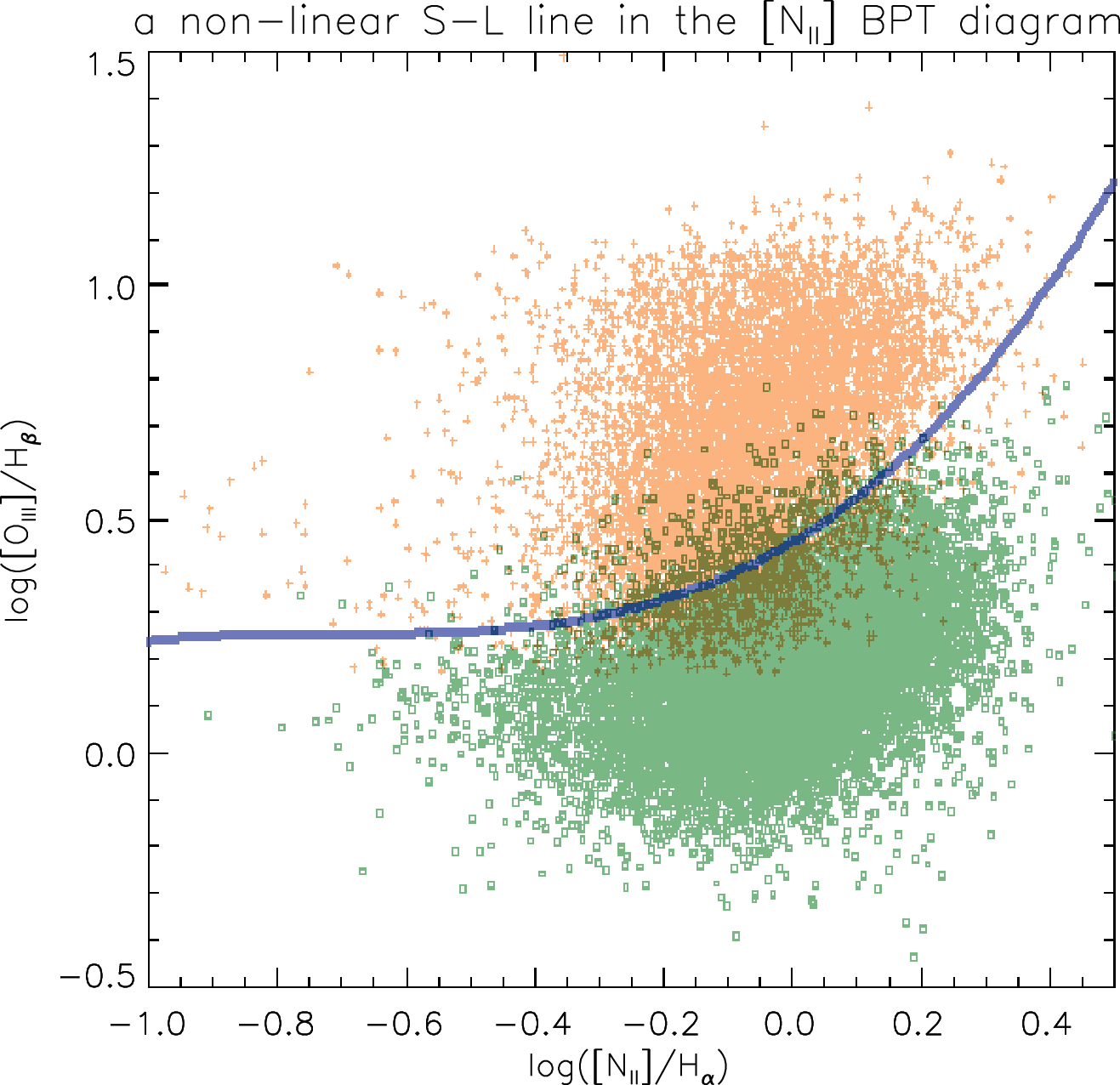}
\caption{This figure demonstrates a semi-transparent visual presentation of the intersection boundary between Seyferts and LINERs in 
the [N~{\sc ii}] BPT diagram. The non-linear line in blue represents a simple outline of the intersection boundary determined by 
visual inspection. The plus signs in orange and hollow squares in dark green represent the Seyferts and LINERs which have been 
classified in both the [S~{\sc ii}] and [O~{\sc i}] BPT diagrams, respectively.}
\label{fig:nii}
\end{figure}

	Meanwhile, \citet{2007MNRAS.382.1415S, 2010MNRAS.403.1036C} have proposed S-L lines in the [N~{\sc ii}] BPT diagram, which 
limited by their methods, can only be linear diagonal lines anyway. However, based on the Ke01 lines and the S-L lines reported in the 
[S~{\sc ii}] and [O~{\sc i}] BPT diagrams \citep{2001ApJ...556..121K,2006MNRAS.372..961K}, 5,791 Seyferts and 4,228 LINERs, among our 
collected narrow emission line galaxies, that are consistently classified in both the [S~{\sc ii}] and [O~{\sc i}] diagrams 
are collected. This approach ensures that only galaxies with unambiguous Seyfert or LINER classifications are used for 
further analysis in the [N~{\sc ii}] diagram. These galaxies are then visualized in the [N~{\sc ii}] BPT diagram, as shown by the 
semi-transparent results in Fig.~\ref{fig:nii}. Visually, they outline a non-linear S-L line between Seyferts and LINERs, with the 
potential to further classify Seyferts and LINERs in the composite galaxy region.

	Therefore, at the current stage, the widely applied method in \citet{2006MNRAS.372..961K} cannot lead to an accepted S-L line 
in the [N~{\sc ii}] BPT diagram, and the proposed Sc07 line and Fe10 line should be improved from linear diagonal lines to non-linear 
lines. Here, in the manuscript, through the Seyferts and LINERs classified in the [S~{\sc ii}] and/or [O~{\sc i}] BPT diagrams 
visualized in the [N~{\sc ii}] BPT diagram, the expected intersection boundary between these two sub-classes of galaxies will 
be considered as the new and improved S-L line in the [N~{\sc ii}] BPT diagram.

\subsection{Our S-L results in the [N~{\sc ii}] BPT diagram}  \label{result}

\begin{figure*}
	\centering
	\includegraphics[width=2\columnwidth]{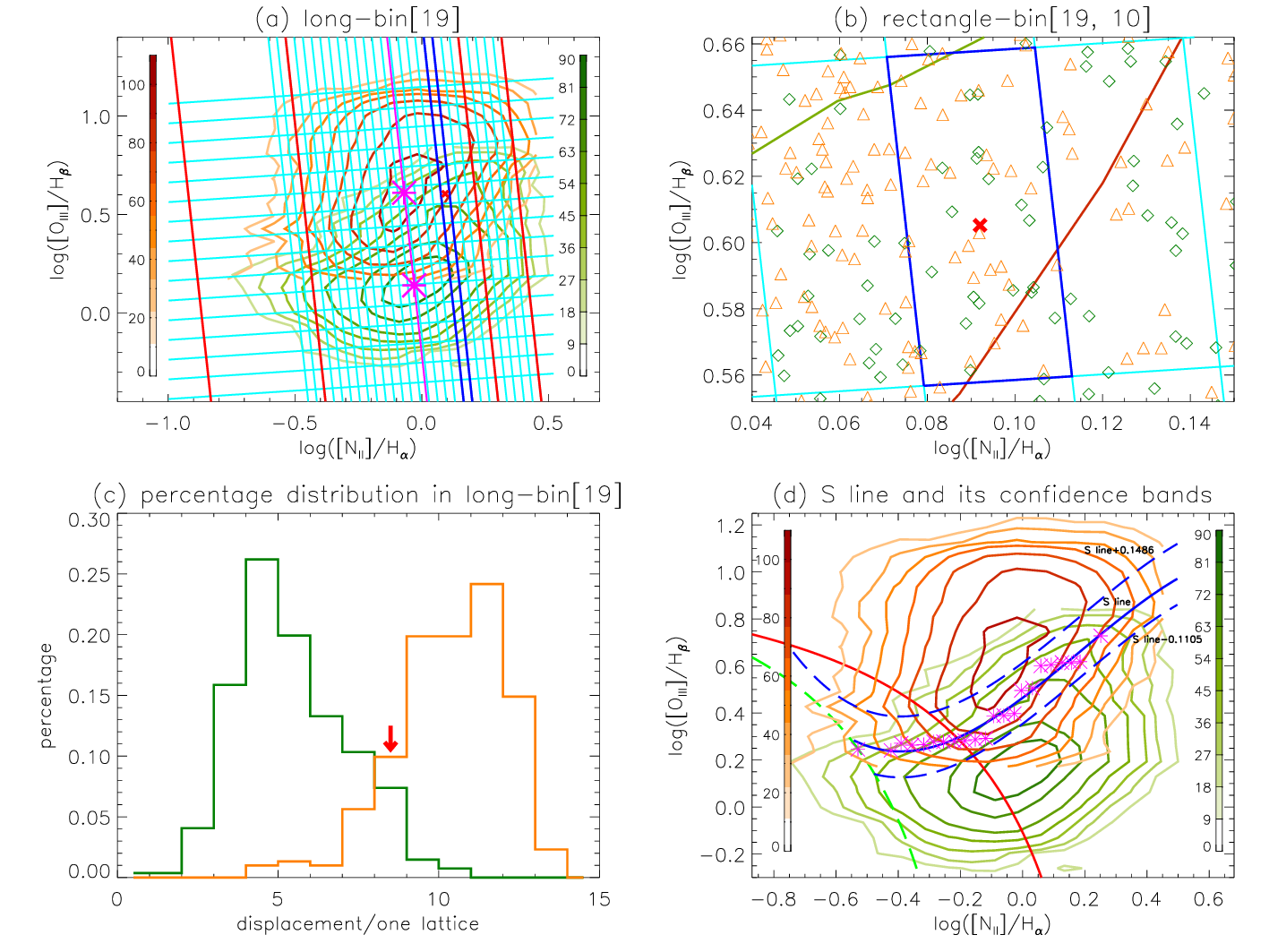}
	\caption{This figure demonstrates the complete process for defining the S line. (a) shows the long-bin[19]. The contours 
	with levels shown with colors from the color table of RED TEMPERATURE and from the color table of green/white LINEAR 
	represent the results for the classified Seyferts and LINERs, respectively. The corresponding number densities of the 
	contour levels are shown in the colorbars on both sides of the panel. The huge magenta asterisks represent the central 
	points of the two contours. The magenta line represents the baseline. Two sets of cyan parallel lines divide the contour 
	maps into 27 $\times$ 15 rectangle-bins. The boundaries of long-bin[19] are highlighted in blue. The boundaries of wider 
	long-bins are highlighted in red. The red cross represents the intersection point of the Seyferts and LINERs distributions 
	within the long-bin[19]. (b) shows the rectangle-bin[19, 9]. The boundaries of rectangle-bin[19, 9] are highlighted in blue. 
	The triangles in orange and the diamonds in dark green represent the Seyferts and LINERs classified in the [S~{\sc ii}] 
	BPT diagram, respectively. The red cross has the same meaning as that in panel (a). (c) shows the couple of histograms 
	depicting the dependence of the percentage of Seyferts (orange) and LINERs (dark green) counts within each rectangle-bin 
	in long-bin[19] relative to their total counts in the long-bin[19] on the displacement of each rectangle-bin in long-bin[19] 
	from the bottom of the long-bin[19]. The red arrow corresponds to the intersection point in long-bin[19], which matches 
	the location of the thick red crosses in the previous two panels. (d) shows the final result of the S line (solid blue line) 
	and its confidence bands (dashed blue lines). Magenta asterisks represent the collected intersection points from all the 
	long-bins and wider long-bins. The solid red line is the Ke01 line. The dashed green line is the Ka03 line.}
	\label{fig:s-bin}
\end{figure*}

\begin{figure*}
	\centering
	\includegraphics[width=2\columnwidth]{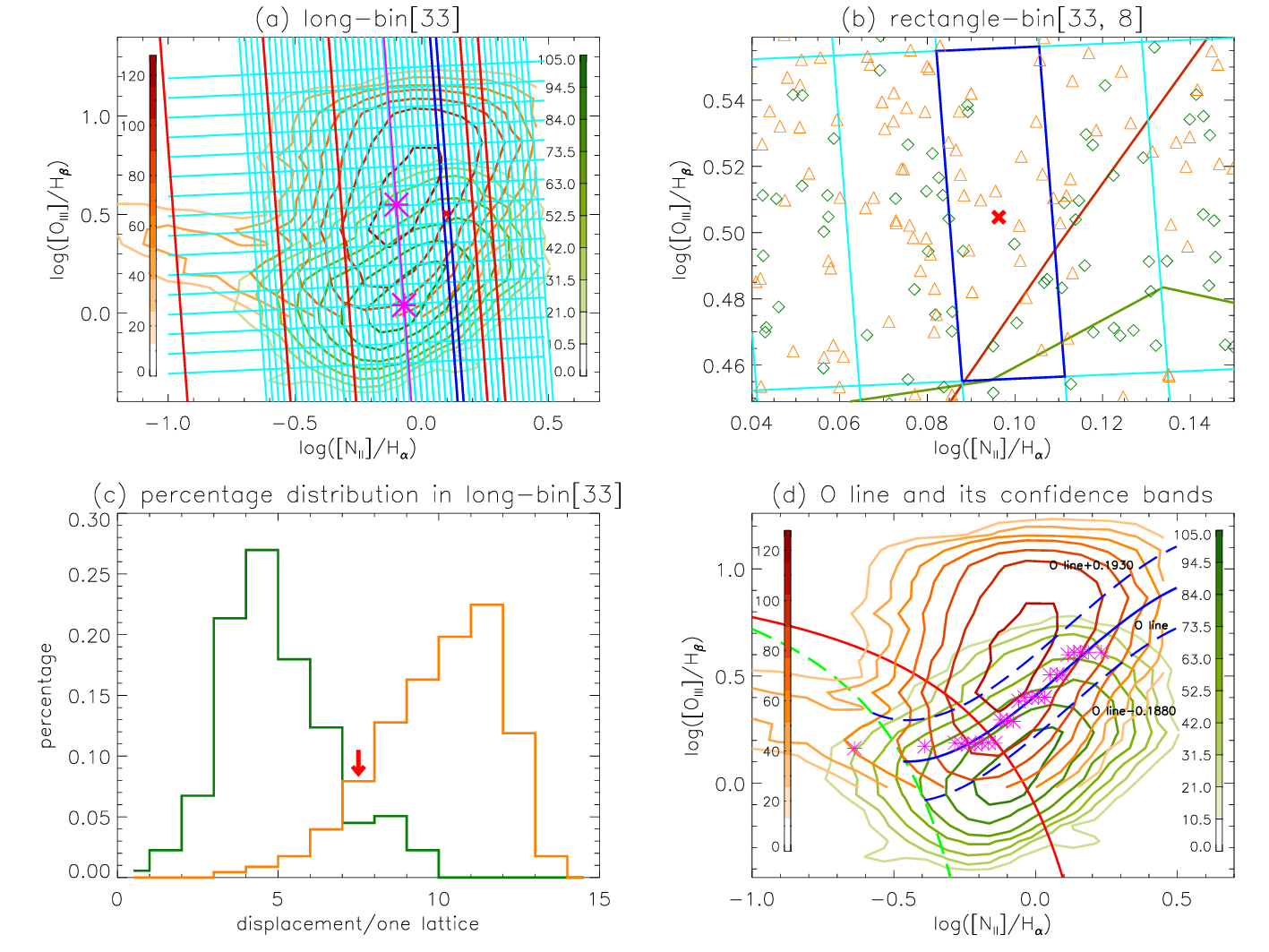}
	\caption{This figure demonstrates the complete process for defining the O line. The symbols and line 
    styles have the same meanings as those in Fig.~\ref{fig:s-bin}, but Seyferts and LINERs are classified in
    the [O~{\sc i}] BPT diagram.}
	\label{fig:o-bin}
\end{figure*}

\begin{figure*}
	\centering
	\includegraphics[width=2\columnwidth]{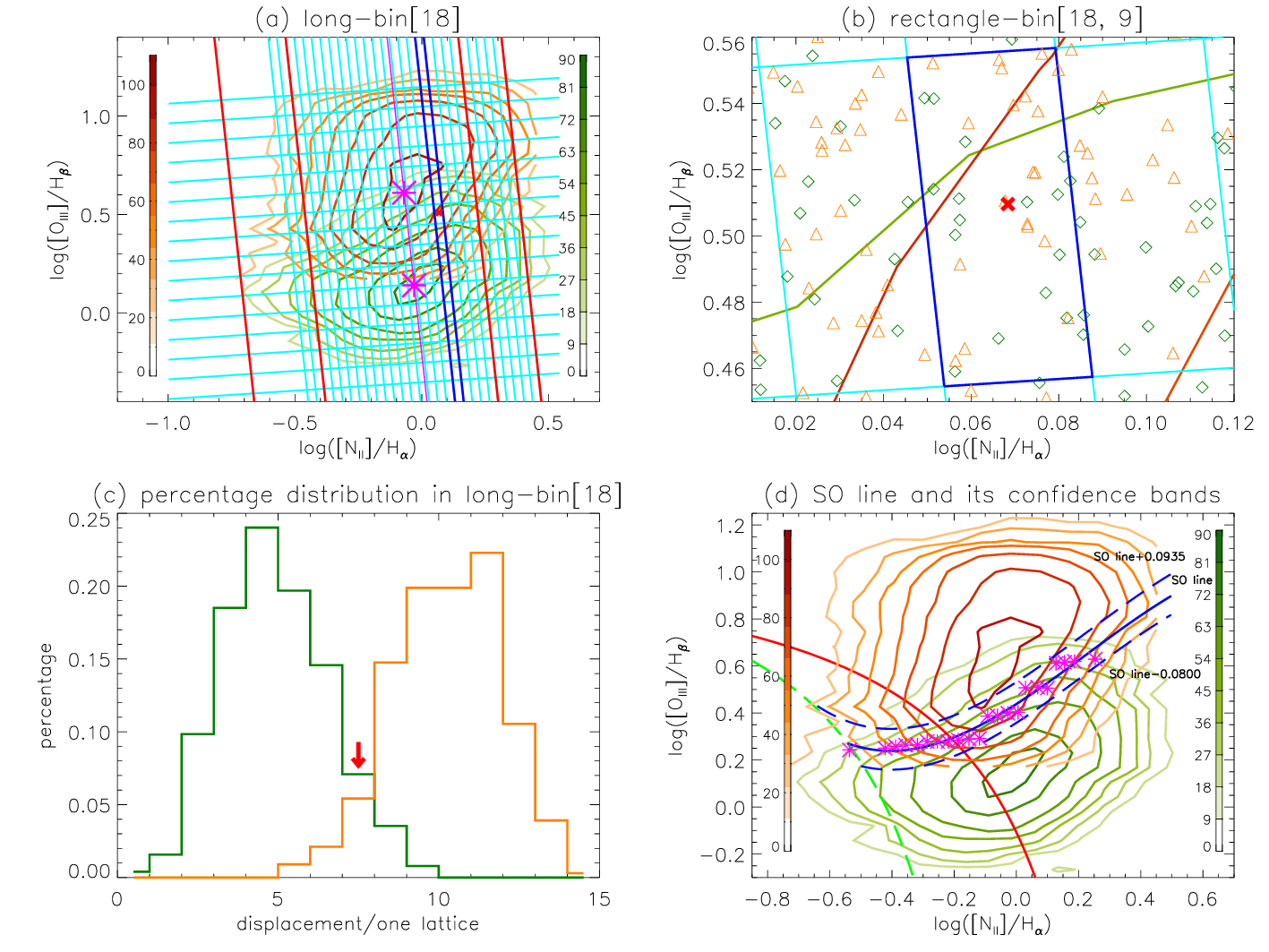}
	\caption{This figure demonstrates the complete process for defining the SO line. The symbols and line 
    styles have the same meanings as those in Fig.~\ref{fig:s-bin}, but Seyferts and LINERs are classified in
    both the [S~{\sc ii}] BPT diagram and the [O~{\sc i}] BPT diagram.}
	\label{fig:so-bin}
\end{figure*}

	Based on the classifications reported in the \citet{2006MNRAS.372..961K}, the Seyferts and LINERs classified only in the 
[S~{\sc ii}] BPT diagram, only in the [O~{\sc i}] diagram, in both the [S~{\sc ii}] BPT diagram and the [O~{\sc i}] BPT diagram are 
visualized in the [N~{\sc ii}] BPT diagram. Then, the one with the highest and most harmonious classification efficiency among the 
corresponding intersection boundaries of the Seyferts and the LINERs can be determined as the S-L line in the [N~{\sc ii}] BPT diagram.

	First of all, based on the 5,966 Seyferts and 4,919 LINERs classified in the [S~{\sc ii}] BPT diagram well represented in 
the [N~{\sc ii}] BPT diagram as the two contour maps shown in panel (a) of Fig.~\ref{fig:s-bin}, the ridge line (if it exists) of 
their contour can be determined as the S-L line in the [N~{\sc ii}] BPT diagram (the S line) by the following steps.

	For the first step, a magenta baseline is drawn to connect the central points of the contour maps for Seyferts and LINERs, 
marked by huge magenta asterisks at (-0.03, 0.14) and (-0.07, 0.61), respectively, as shown in panel (a) of Fig.~\ref{fig:s-bin}. 
For the second step, 28 parallel lines are drawn along the direction of the baseline, with each line spaced 0.04 dex apart, 
dividing the contour maps for Seyferts and LINERs into 27 long-bins, as shown by an example long-bin[19] in panel (a) of 
Fig.~\ref{fig:s-bin}. For the third step, 16 parallel lines are drawn in the direction perpendicular to the baseline, with each 
line spaced 0.09 dex apart, further subdividing each long-bin into 15 rectangle-bins, as shown by an example rectangle-bin[19, 10] 
in panel (b) of Fig.~\ref{fig:s-bin}. The contour maps for Seyferts and LINERs are divided into 27 $\times$ 15 rectangle-bins, as 
shown in the grids in panel (a) of Fig.~\ref{fig:s-bin}. Meanwhile, the number counts and average coordinates of the classified Seyferts 
and LINERs within each rectangle-bin can be easily obtained. In fact, grids denser or sparser than 28 $\times$ 16 have been tested, 
yielding consistently similar results, except when the grids are extremely large or small, which results in either too few or too 
many Seyferts and LINERs located within the rectangle-bins. For the fourth step, a couple of histograms are drawn to depict the 
dependence of the percentages of Seyferts and LINERs counts within each rectangle-bin relative to their total counts in the 
corresponding long-bin on the displacement of each rectangle-bin from the bottom of the corresponding long-bin. A couple of example 
histograms are shown in panel (c) of Fig.~\ref{fig:s-bin}. The average of the average coordinates of Seyferts and LINERs in the 
rectangle-bin corresponding to the local minimum point between the peaks of the two histograms is considered as the coordinates 
of the intersection point in each long-bin. In fact, due to the small sample sizes of Seyferts and LINERs, reasonable intersection 
points cannot be pinpointed from long-bin[1] to long-bin[2] and from long-bin[23] to long-bin[27]. In order to better pinpoint the 
intersection points near log([N~{\sc ii}]/H$\alpha$)=0.3 and -0.5, two wider long-bins are established, with boundaries marked in 
red in panel (a) of Fig.~\ref{fig:s-bin}. Finally, the coordinates of the intersection points in all the long-bins and wider 
long-bins are pinpointed. Then, considering the non-linear trend of the intersection points, a third-degree polynomial function, 
instead of a straight line, is applied using the poly\_fit code without considering uncertainties to describe the intersection 
points, leading to the S line, which determines the intersection boundary between Seyferts and LINERs in the [N~{\sc ii}] BPT diagram, 
as shown in panel (d) of Fig.~\ref{fig:s-bin}, by the formula:
\begin{equation}
\begin{aligned}
	\log&(\text{\oiii}) = 0.46 + 0.99 \times \log(\text{\nii}) \\
	    &+ 0.61 \times (\log(\text{\nii}))^2 - 1.11 \times (\log(\text{\nii}))^3
\end{aligned}
\end{equation}.
Moreover, among the 21,613 galaxies above the Ka03 line in the [N~{\sc ii}] BPT diagram, 6,301 galaxies are classified as Seyferts and 
15,312 as LINERs by the S line. In contrast, 7,424 galaxies are classified as Seyferts and 14,189 as LINERs by the Sc07 line, while 
6,815 galaxies are classified as Seyferts and 14,798 as LINERs by the Fe10 line. Meanwhile, among the 9,797 galaxies above the Ke01 line 
in the [N~{\sc ii}] BPT diagram, 5,866 galaxies are classified as Seyferts and 3,931 as LINERs by the S line. In contrast, 6,107 galaxies 
are classified as Seyferts and 3,690 as LINERs by the Sc07 line, while 5,838 galaxies are classified as Seyferts and 3,959 as LINERs by 
the Fe10 line.

	Similar to the S line determined above in the [N~{\sc ii}] BPT diagram through the classified Seyferts and LINERs in 
the [S~{\sc ii}] BPT diagram, another S-L line (the O line) in the [N~{\sc ii}] BPT diagram can be determined through the 
classified Seyferts and LINERs in the [O~{\sc i}] BPT diagram as follows.

        The 7,550 Seyferts and 5,620 LINERs classified in the [O~{\sc i}] BPT diagram are represented in the [N~{\sc ii}] BPT 
diagram, and the aforementioned steps are repeated to determine the corresponding intersection boundary, referred to as the O line. 
The central points of the contour maps for Seyferts and LINERs are at coordinates (-0.10, 0.55) and (-0.07, 0.04). It is worth 
noting that when drawing parallel lines along the baseline direction, the spacing is adjusted to 0.02 dex, otherwise, the numbers 
of galaxies in the rectangle-bins will be too large. Moreover, three wider long-bins are established near log([N~{\sc ii}]/H$\alpha$)=0.2, 
-0.4 and -0.6, due to the existence of a significant number of Seyferts and LINERs. The detailed process is shown in Fig.~\ref{fig:o-bin}, 
with the determined O line described by the formula:
\begin{equation}
\begin{aligned}
	\log&(\text{\oiii}) = 0.41 + 1.11 \times \log(\text{\nii}) \\
	    &+ 0.40 \times (\log(\text{\nii}))^2 - 1.22 \times (\log(\text{\nii}))^3
\end{aligned}
\end{equation}.
Similarly, among the 21,613 galaxies located above the Ka03 line in the [N~{\sc ii}] BPT diagram, 7,624 galaxies are classified as Seyferts 
and 13,989 as LINERs by the O line. Meanwhile, among the 9,797 galaxies located above the Ke01 line in the [N~{\sc ii}] BPT diagram, 6,405 
galaxies are classified as Seyferts and 3,392 as LINERs by the O line. As shown in panel (d) of Fig.~\ref{fig:o-bin}, the O line has been 
presented in the [N~{\sc ii}] BPT diagram.

	Finally, similar to the S line and the O line determined above in the [N~{\sc ii}] BPT diagram, another S-L line 
(the SO line) in the [N~{\sc ii}] BPT diagram can be determined through the classified Seyferts and LINERs both in the 
[S~{\sc ii}] BPT diagram and in the [O~{\sc i}] BPT diagram as follows.

	The 5,791 Seyferts and 4,228 LINERs classified in both the [S~{\sc ii}] and the [O~{\sc i}] BPT diagram are 
represented in the [N~{\sc ii}] BPT diagram. The central points of the contour maps for Seyferts and LINERs are at coordinates 
(-0.03, 0.14) and (-0.07, 0.61). By repeating the aforementioned steps, with the detailed process shown in Fig.~\ref{fig:so-bin}, 
a third S-L line is determined, referred to as the SO line, with the equation:
\begin{equation}
\begin{aligned}
	\log&(\text{\oiii}) = 0.44 + 0.85 \times \log(\text{\nii}) \\
	    &+ 0.55 \times (\log(\text{\nii}))^2 - 0.82 \times (\log(\text{\nii}))^3
\end{aligned}
\end{equation}.
Similarly, among the 21,613 galaxies located above the Ka03 line in the [N~{\sc ii}] BPT diagram, 6,493 galaxies are classified 
as Seyferts and 15,120 as LINERs by the SO line. Meanwhile, among the 9,797 galaxies located above the Ke01 line in the [N~{\sc ii}] 
BPT diagram, 6,067 galaxies are classified as Seyferts and 3,730 as LINERs by the SO line. As shown in panel (d) of 
Fig.~\ref{fig:so-bin}, the SO line has been presented in the [N~{\sc ii}] BPT diagram.

\begin{figure*}[ht!]
\centering
\includegraphics[width=2\columnwidth]{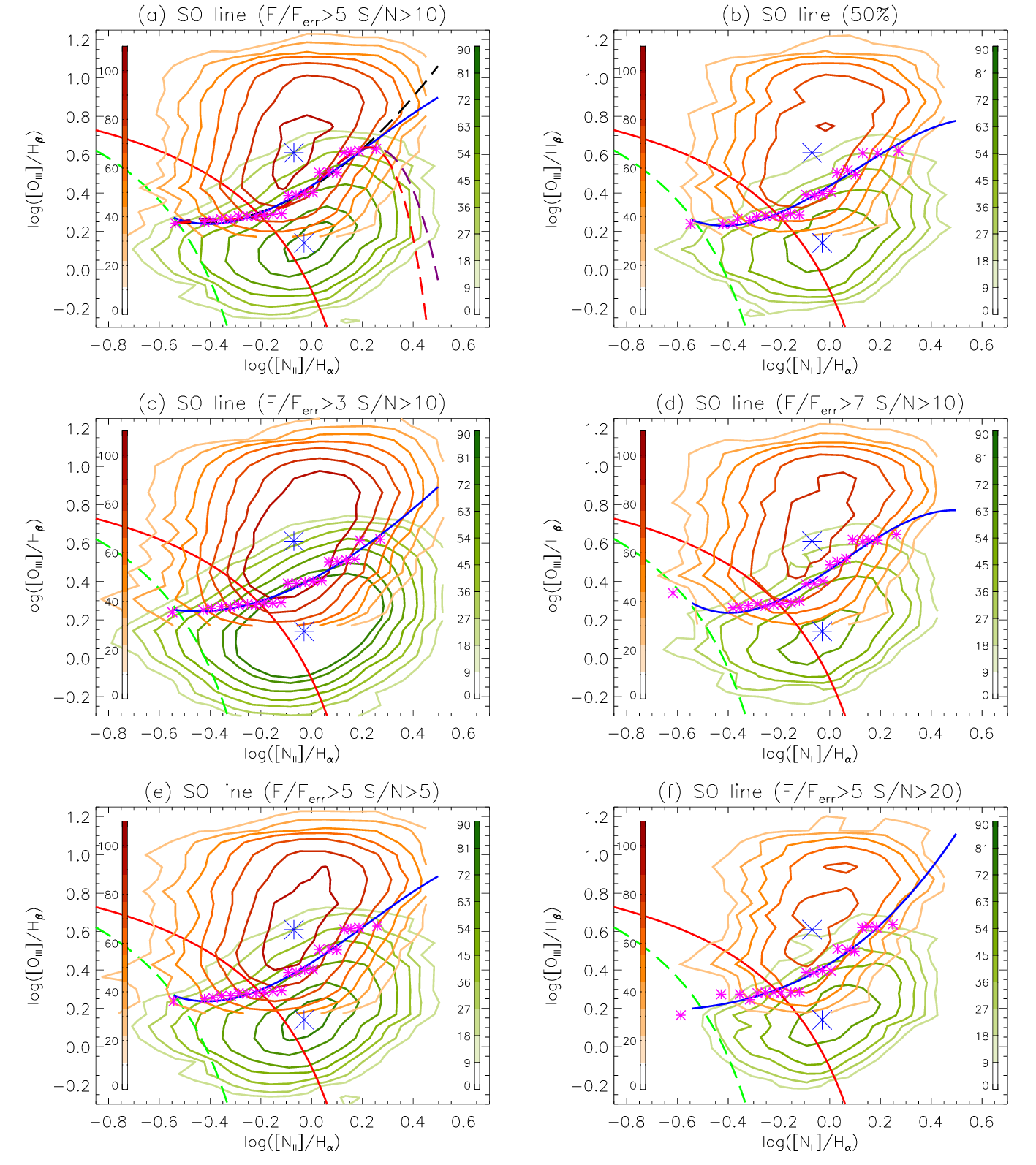}
\caption{This figure shows the results of determining the SO line through the same procedure using different samples and the 
results of fitting the SO line using polynomial functions from 2nd to 5th degrees. Panel (a) presents the results of determining the SO 
line using sample 1 with polynomial functions from 2nd to 5th degrees. The symbols and line styles have the same meanings as those in 
Fig.~\ref{fig:s-bin}. The large blue asterisks represent the central points of the two contours. The dashed black curve represents the 
2nd-degree fit. The solid blue curve represents the 3rd-degree fit. The dashed purple curve represents the 4th-degree fit. The dashed red 
curve represents the 5th-degree fit. Panels (b), (c), (d), (e) and (f) present the results of determining the SO line using samples 2 to 
6 with a 3rd-degree fit, respectively. The symbols and line styles have the same meanings as those in panel (a).}
\label{fig:representative}
\end{figure*}

\begin{table*}
\centering
\caption{\centering Samples Corresponding to Various SQL Conditions}
\begin{tabular}{lccccc}
\hline
\noalign{\smallskip}
Sample Number & Fluxes/Uncertainties & SNmedian & Total Number & Counts of Seyferts & Counts of LINERs\\   
\noalign{\smallskip}
\hline
\noalign{\smallskip}
Sample 1    & 5    & 10  & 47,968 & 5,791 & 4,228  \\
Sample 2    & 5    & 10  & 23,984 & 2,887 & 2,115  \\
Sample 3    & 3    & 10  & 95,364 & 8,474 & 11,445 \\
Sample 4    & 7    & 10  & 26,239 & 3,928 & 2,019  \\   
Sample 5    & 5    & 5   & 53,344 & 6,199 & 4,395  \\   
Sample 6    & 5    & 20  & 16,194 & 2,065 & 2,570  \\   
\noalign{\smallskip}
\hline
\noalign{\smallskip}
\end{tabular}
\tablefoot{This table shows different samples of narrow emission line galaxies collected using various SQL conditions.
Column 1 is the sample number. Columns 2 and 3 are the lower limits of the flux-to-uncertainty ratio and $SNmedian$ for the emission lines 
of the narrow emission line galaxies. Column 4 is the total number of narrow emission line galaxies in each sample. Columns 5 and 6 are the 
counts of galaxies consistently classified as Seyferts and LINERs in both the [S~{\sc ii}] and [O~{\sc i}] BPT diagrams, respectively.}
\label{tab:sam}
\end{table*}

There are the following points that need to be declared.
\begin{enumerate}
    \item First, taking the SO line as an example, the above procedures are repeated with samples at different scales to examine 
the impacts of selection effects, as well as the robustness of our method. 50$\%$ of the objects are randomly selected from Sample 1 as one new 
sample (Sample 2), and the other four samples (Samples 3-6) are constructed by adjusting the lower limits of the flux-to-uncertainty ratio or 
$SNmedian$ for the fluxes in the SQL Search query. These operations lead to five new couples of Seyferts and LINERs samples. Detailed SQL 
condition adjustments and counts of Seyferts and LINERs are presented in Table~\ref{tab:sam}. Subsequently, the same procedures used to determine 
the SO line (as shown in panel (a) of Fig.~\ref{fig:representative}) are repeated, leading to five classification lines (as shown in panels (b), 
(c), (d), (e) and (f) of Fig.~\ref{fig:representative}). It is evident that the intersection points in panels (b), (c), (d), (e) and (f) show no 
significant differences in position and trend compared to those in panel (a). This consistency indicates that, while different SQL search conditions 
introduce certain selection effects, their impacts on determining the SO line are negligible. Therefore, our method for classifying Seyferts 
and LINERs in the [N~{\sc ii}] BPT diagram can be considered highly robust.
    \item Second, multiple polynomial forms, from 2nd-degree to 5th-degree, have been systematically evaluated during the line fitting. 
Although the 4th-degree and 5th-degree polynomial fits exhibit smaller RMS scatter and $\chi^2/\text{dof}$ values, as shown in Table~\ref{tab:polyfit}, 
they show unphysically sharp declines where $\log([N_{\text{II}}]/H\alpha)$ > 0.3, while the 2nd-degree polynomial fit shows an unphysically sharp 
increase in the same region, as shown in panel (a) of Fig.~\ref{fig:representative}. In the region with a high density of intersection points, all 
fits perform comparably. However, although polynomial fits of 2nd, 4th and 5th degrees may be valid within limited regimes, the 3rd-degree polynomial 
offers the best compromise between smoothness, physical plausibility and performance across the entire parameter space, making it the most suitable choice.

\begin{table}[ht!]
\centering
\caption{\centering RMS scatters and $\chi^2/\text{dof}$ values for 2nd-5th Degree Polynomial Fits}
\label{tab:polyfit}
\begin{tabular}{lcc}
\hline
\noalign{\smallskip}
Degree & RMS scatters & $\chi^2/\text{dof}$ values\\
\noalign{\smallskip}
\hline
\noalign{\smallskip}
2nd    & 0.0309867    & 10.6124     \\
3rd    & 0.0296985    & 10.2900     \\
4th    & 0.0223435    & 6.16696     \\
5th    & 0.0216513    & 6.15270     \\   
\noalign{\smallskip}
\hline
\noalign{\smallskip}
\end{tabular}
\tablefoot{This table shows the RMS scatters and $\chi^2/\text{dof}$ values fitted with a polynomial from the 2nd to the 5th degree 
when fitting the SO line. Column 1 is the highest degree. Column 2 is the RMS scatters. Column 3 is the $\chi^2/\text{dof}$ values.}
\end{table}

    \item Third, the orientations of the baselines have been adjusted several times by modifying the coordinates of the central 
points in the contour maps for Seyferts and LINERs, and the resulting intersection boundaries, obtained through repeated adjustments, 
show significant overlap, confirming the reliability of the results.  
\end{enumerate}
So far, we have five S-L lines in the [N~{\sc ii}] BPT diagram, the S line, the O line, the SO line, the Sc07 line and the Fe10 line. 
It is necessary to check which one is preferred.

\section{Necessary Discussions} \label{sec:discussions}

	In order to test the efficiency of the S line, the O line, the SO line, the Sc07 line and the Fe10 line, quantitative 
analyses are carried out as follows.

\subsection{Comparisons of number ratios $F_{S}$ and $F_{L}$}

	For the first, the effectiveness of the S-L lines can be checked by the defined number ratios $F_{S}$ and $F_{L}$, which 
are defined by
\begin{equation}
	\begin{aligned}
		&F_{S}= N_{S_{a}} / N_{S_{t}}\\
		&F_{L}= N_{L_{b}} / N_{L_{t}}\\
	\end{aligned}
\end{equation}.
$N_{S_{t}}$ ($N_{L_{t}}$) is the total number of Seyferts (LINERs) classified in the [S~{\sc ii}] and/or [O~{\sc i}] BPT diagrams 
and lying above the Ka03 (Ke01) line in the [N~{\sc ii}] BPT diagram. $N_{S_{a}}$ ($N_{L_{b}}$) is the number of Seyferts (LINERs) 
classified in the [S~{\sc ii}] and/or [O~{\sc i}] BPT diagrams and lying above the Ka03 (Ke01) line and also above (below) the 
given S-L line in the [N~{\sc ii}] BPT diagram. A larger $F_{S}$ ($F_{L}$) demonstrates that more Seyferts (LINERs) are classified 
into the same category through the given S-L line, indicating higher classification efficiency. Meanwhile, a smaller absolute 
difference of $|F_{S} - F_{L}|$ for the S-L line indicates a more harmonious consistency in the classification efficiency of 
Seyferts and LINERs. The $F_{S}$, $F_{L}$ and corresponding $|F_{S} - F_{L}|$ for the S line, the O line, the SO line, the Sc07 
line and the Fe10 line are shown in Table~\ref{tab:F}, with detailed descriptions on the results as follows.

\begin{table*}
    \centering
    \caption{\centering Number Ratios $F_{S}$ and $F_{L}$ of S-L Lines}
	 \begin{tabular}{lcccccccc}
		\hline
      \noalign{\smallskip}
		          & $N_{S_{t}}$ & $N_{S_{a}}$ & $F_{S}$ ($\%$) & $N_{L_{t}}$ & $N_{L_{b}}$ & $F_{L}$ ($\%$) & $|F_{S} - F_{L}|$ \\
		\noalign{\smallskip}
      \hline
      \noalign{\smallskip}
      \multicolumn{8}{c}{Ka03} \\
      \noalign{\smallskip}
		 S line    & 5,927        & 5,440        & 91.78          & 4,848        & 4,438        & 91.54          & 0.24          \\
		 O line    & 7,357        & 6,353        & 86.35          & 5,462        & 4,857        & 88.92          & 2.57          \\
		 SO line   & 5,754        & 5,405        & 93.93          & 4,170        & 3,945        & 94.60          & 0.67          \\
		 Sc07 line & 5,754        & 5,442        & 94.58          & 4,170        & 3,758        & 90.12          & 4.46          \\
		 Fe10 line & 5,754        & 5,328        & 92.60          & 4,170        & 3,870        & 92.81          & 0.21          \\
		 \noalign{\smallskip}
      \hline
      \noalign{\smallskip}
      \multicolumn{8}{c}{Ke01} \\
      \noalign{\smallskip}
      S line    & 5,711        & 5,275        & 92.37          & 3,368        & 3,004        & 89.19          & 3.17          \\
      O line    & 6,544        & 5,911        & 90.33          & 3,044        & 2,670        & 87.71          & 2.61          \\
      SO line   & 5,574        & 5,273        & 94.60          & 2,814        & 2,623        & 93.21          & 1.39          \\
      Sc07 line & 5,574        & 5,272        & 94.58          & 2,814        & 2,622        & 93.18          & 1.41          \\
      Fe10 line & 5,574        & 5,167        & 92.70          & 2,814        & 2,677        & 95.13          & 2.43          \\
      \noalign{\smallskip}
      \hline
      \noalign{\smallskip}
	 \end{tabular}
    \tablefoot{This table demonstrates the values for number ratios $F_{S}$ and $F_{L}$ of each S-L line. 
The upper (lower) part of the table describes the calculation results of $F_{S}$ and $F_{L}$ based on galaxies 
lying above the Ka03 (Ke01) line in the [N~{\sc ii}] BPT diagram.
         Column 1 shows the different S-L lines. 
	      Column 2 shows the $N_{S_{t}}$ for each S-L line.
	      Column 3 shows the $N_{S_{a}}$ for each S-L line. 
	      Column 4 shows the $F_{S}$ for each S-L line, with the results in percentage form.
	      Column 5 shows the $N_{L_{t}}$ for each S-L line.
	      Column 6 shows the $N_{L_{b}}$ for each S-L line.
	      Column 7 shows the $F_{L}$ for each S-L line, with the results in percentage form.
	      Column 8 shows the $|F_{S} - F_{L}|$ for each S-L line.}
\label{tab:F}
\end{table*}

\begin{itemize}
	\item For the S line, Seyferts and LINERs are classified in the [S~{\sc ii}] BPT diagram, with $N_{S_{t}}$ = 5,927, 
	$N_{L_{t}}$ = 4,848, $N_{S_{a}}$ = 5,440 and $N_{L_{b}}$ = 4,438, based on the Ka03 line. 
  Then, $F_{S} = N_{S_{a}} / N_{S_{t}}$ = 91.78$\%$ and $F_{L} = N_{L_{b}} / N_{L_{t}}$ = 91.54$\%$, with an absolute 
  difference of $|F_{S} - F_{L}|$ = 0.24$\%$.
	\item For the O line, Seyferts and LINERs are classified in the [O~{\sc i}] BPT diagram, with $N_{S_{t}}$ = 7,357, 
	$N_{L_{t}}$ = 5,462, $N_{S_{a}}$ = 6,353 and $N_{L_{b}}$ = 4,857, based on the Ka03 line. 
  Then, $F_{S} = N_{S_{a}} / N_{S_{t}}$ = 86.35$\%$ and $F_{L} = N_{L_{b}} / N_{L_{t}}$ = 88.92$\%$, with an absolute 
  difference of $|F_{S} - F_{L}|$ = 2.57$\%$.
	\item For the SO line, Seyferts and LINERs are classified in the [S~{\sc ii}] and [O~{\sc i}] BPT diagram, with 
	$N_{S_{t}}$ = 5,754, $N_{L_{t}}$ = 4,170, $N_{S_{a}}$ = 5,405 and $N_{L_{b}}$ = 3,945, based on the Ka03 line. Then, 
	$F_{S} = N_{S_{a}} / N_{S_{t}}$ = 93.93$\%$ and $F_{L} = N_{L_{b}} / N_{L_{t}}$ = 94.60$\%$, with an absolute difference 
	of $|F_{S} - F_{L}|$ = 0.67$\%$.
	The $F_{S}$ and $F_{L}$ for the SO line are the largest among the results for the S line, the O line and the SO line, 
	with a small enough absolute difference of $|F_{S} - F_{L}|$.
	\item For the Sc07 line, Seyferts and LINERs are classified in both the [S~{\sc ii}] diagram and the [O~{\sc i}] BPT diagram, 
	with $N_{S_{t}}$ = 5,754, $N_{L_{t}}$ = 4,170, $N_{S_{a}}$ = 5,442 and $N_{L_{b}}$ = 3,758, based on the Ka03 line. Then, 
	$F_{S} = N_{S_{a}} / N_{S_{t}}$ = 94.58$\%$ and $F_{L} = N_{L_{b}} / N_{L_{t}}$ = 90.12$\%$, with an absolute difference 
	of $|F_{S} - F_{L}|$ = 4.46$\%$. Although $F_{S}$ = 94.58$\%$ is 0.65$\%$ larger than that for the SO line, the 
   absolute difference for the Sc07 line is 3.79$\%$ larger than that for the SO line.
	\item For the Fe10 line, Seyferts and LINERs are classified in the [S~{\sc ii}] and [O~{\sc i}] BPT diagram, with 
	$N_{S_{t}}$ = 5,754, $N_{L_{t}}$ = 4,170, $N_{S_{a}}$ = 5,328 and $N_{L_{b}}$ = 3,870, based on the Ka03 line. Then, 
	$F_{S} = N_{S_{a}} / N_{S_{t}}$ = 92.60$\%$ and $F_{L} = N_{L_{b}} / N_{L_{t}}$ = 92.81$\%$, with an absolute difference 
	of $|F_{S} - F_{L}|$ = 0.21$\%$. Although the absolute difference for the Fe10 line is the smallest among all the S-L lines, 
   the $F_{S}$ and $F_{L}$ for the Fe10 line are 1.33$\%$ and 1.79$\%$ smaller than those for the SO line, respectively.
  \item For the S line, Seyferts and LINERs are classified in the [S~{\sc ii}] BPT diagram, with $N_{S_{t}}$ = 5,711, 
	$N_{L_{t}}$ = 3,368, $N_{S_{a}}$ = 5,275 and $N_{L_{b}}$ = 3,004, based on the Ke01 line. 
  Then, $F_{S} = N_{S_{a}} / N_{S_{t}}$ = 92.37$\%$ and $F_{L} = N_{L_{b}} / N_{L_{t}}$ = 89.19$\%$, with an absolute 
  difference of $|F_{S} - F_{L}|$ = 3.17$\%$.
	\item For the O line, Seyferts and LINERs are classified in the [O~{\sc i}] BPT diagram, with $N_{S_{t}}$ = 6,544, 
	$N_{L_{t}}$ = 3,044, $N_{S_{a}}$ = 5,911 and $N_{L_{b}}$ = 2,670, based on the Ke01 line. 
  Then, $F_{S} = N_{S_{a}} / N_{S_{t}}$ = 90.33$\%$ and $F_{L} = N_{L_{b}} / N_{L_{t}}$ = 87.71$\%$, with an absolute 
  difference of $|F_{S} - F_{L}|$ = 2.61$\%$.
	\item For the SO line, Seyferts and LINERs are classified in the [S~{\sc ii}] and [O~{\sc i}] BPT diagram, with 
	$N_{S_{t}}$ = 5,574, $N_{L_{t}}$ = 2,814, $N_{S_{a}}$ = 5,273 and $N_{L_{b}}$ = 2,623, based on the Ke01 line. Then, 
	$F_{S} = N_{S_{a}} / N_{S_{t}}$ = 94.60$\%$ and $F_{L} = N_{L_{b}} / N_{L_{t}}$ = 93.21$\%$, with an absolute difference 
	of $|F_{S} - F_{L}|$ = 1.39$\%$.
	The $F_{S}$ and $F_{L}$ for the SO line are the largest among the results for the S line, the O line and the SO line, 
	with the smallest absolute difference of $|F_{S} - F_{L}|$.
	\item For the Sc07 line, Seyferts and LINERs are classified in both the [S~{\sc ii}] diagram and the [O~{\sc i}] BPT diagram, 
	with $N_{S_{t}}$ = 5,574, $N_{L_{t}}$ = 2,814, $N_{S_{a}}$ = 5,272 and $N_{L_{b}}$ = 2,622, based on the Ka03 line. Then, 
	$F_{S} = N_{S_{a}} / N_{S_{t}}$ = 94.58$\%$ and $F_{L} = N_{L_{b}} / N_{L_{t}}$ = 93.18$\%$, with an absolute difference 
	of $|F_{S} - F_{L}|$ = 1.41$\%$. Although $F_{S}$ and $F_{L}$ are the closest to those for the SO line, the 
   absolute difference for the Sc07 line is 0.02$\%$ larger than that for the SO line.
	\item For the Fe10 line, Seyferts and LINERs are classified in the [S~{\sc ii}] and [O~{\sc i}] BPT diagram, with 
	$N_{S_{t}}$ = 5,574, $N_{L_{t}}$ = 2,814, $N_{S_{a}}$ = 5,167 and $N_{L_{b}}$ = 2,677, based on the Ke01 line. Then, 
	$F_{S} = N_{S_{a}} / N_{S_{t}}$ = 92.70$\%$ and $F_{L} = N_{L_{b}} / N_{L_{t}}$ = 95.13$\%$, with an absolute difference 
	of $|F_{S} - F_{L}|$ = 2.43$\%$. Although $F_{L} = 95.13\%$ is 1.92$\%$ larger than that for the SO line, the 
   absolute difference for the Fe10 line is 1.04$\%$ larger than that for the SO line.
\end{itemize}

	Both $F_{S}$ and $F_{L}$ contributed by the SO line are the largest and consistent enough, whether analyzed based on the 
Ka03 line or the Ke01 line, compared to those contributed by the S line and the O line. Considering the SO line, Sc07 line 
and Fe10 line, although neither the $F_{S}$ based on the Ka03 line nor the $F_{S}$ based on the Ke01 line contributed by the 
SO line is the largest, the SO line shows the best balance between efficiency and consistency overall for achieving high and 
consistent $F_{S}$ and $F_{L}$ with a relatively small $|F_{S} - F_{L}|$, making it the most reliable among the three.

\subsection{Comparisons of $F_{S}$ / $F_{L}$}

\begin{figure*}
	\centering
	\includegraphics[width=2\columnwidth]{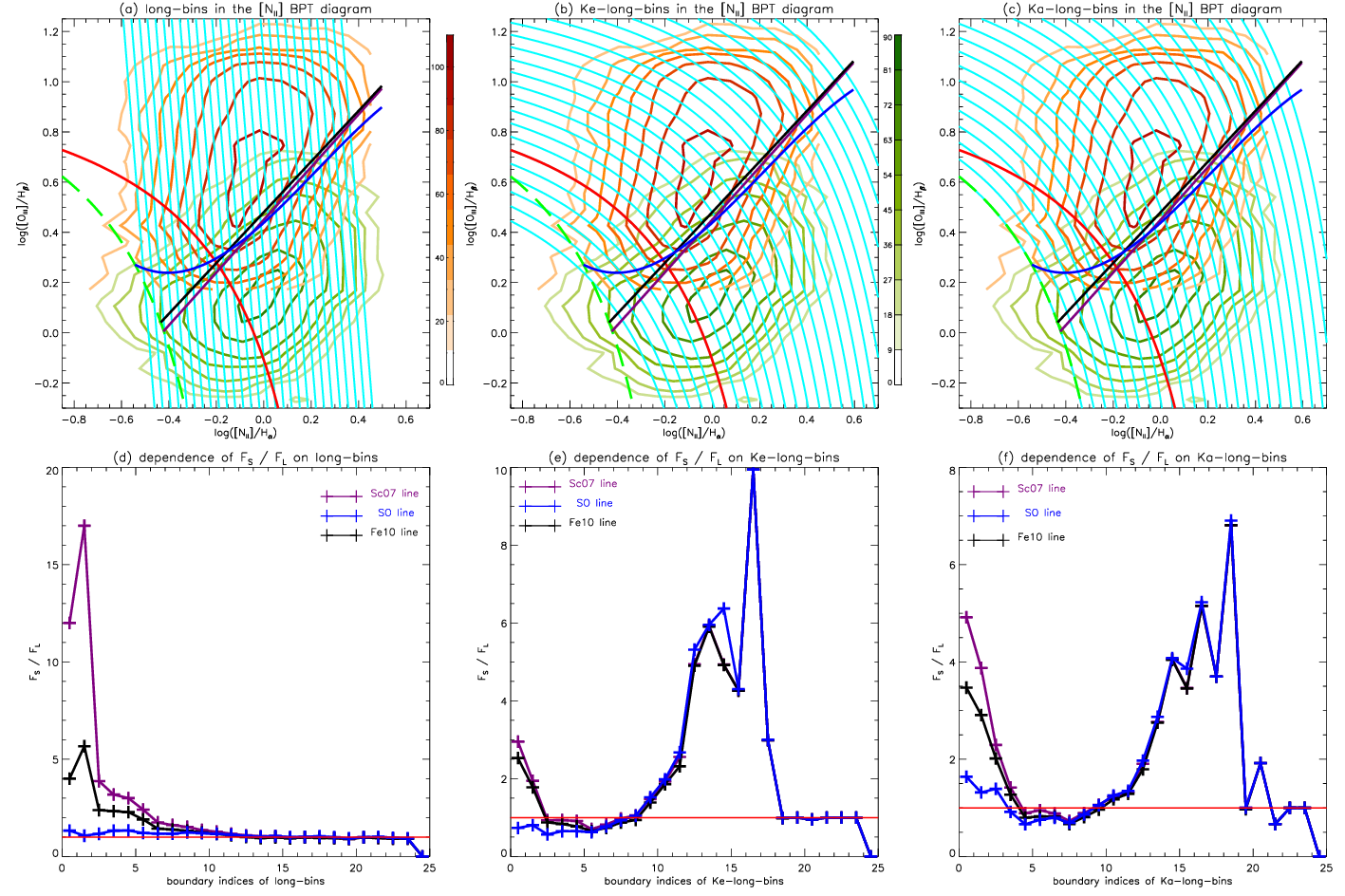}
	\caption{This figure demonstrates the process and results for testing the dependencies of $F_{S}$ / $F_{L}$ on long-bins, 
Ke-long-bins and Ka-long-bins. (a) shows long-bins in the [N~{\sc ii}] BPT diagram. The contours with levels shown with colors 
from the color table of RED TEMPERATURE and from the color table of green/white LINEAR represent the results for the Seyferts 
and LINERs classified in the [S~{\sc ii}] and [O~{\sc i}] BPT diagram, respectively. The corresponding number densities of the 
contour levels are shown in the colorbars on both sides of panel (b). The 28 lines in cyan represent the boundaries of long-bins. 
The solid red line is the Ke01 line. The dashed green line is the Ka03 line. The solid purple line is the Sc07 line. The solid 
black line is the Fe10 line. The solid blue line is the SO line. (b) shows Ke-long-bins in the [N~{\sc ii}] BPT diagram. The 
symbols and line styles have the same meanings as those in panel (a), but the 25 arcs in cyan represent the boundaries of 
Ke-long-bins. (c) shows Ka-long-bins in the [N~{\sc ii}] BPT diagram. The symbols and line styles have the same meanings as 
those in panel (a), but the 25 arcs in cyan represent the boundaries of Ka-long-bins. (d) shows the dependence of $F_{S}$ / $F_{L}$ 
for S-L lines on long-bins. The colors of the polylines correspond to the colors of the S-L lines in panel (a), purple for the Sc07 
line, black for the Fe10 line and blue for the SO line. The red horizontal line indicates log(\nii) = 1. (e) shows the dependence of 
$F_{S}$ / $F_{L}$ for S-L lines on Ke-long-bins. The other symbols and line styles have the same meanings as those in panel (d). 
(f) shows the dependence of $F_{S}$ / $F_{L}$ for S-L lines on Ka-long-bins. The other symbols and line styles have the same meanings 
as those in panel (d).}
\label{fig:fluctuation}
\end{figure*}

       For the second, the ratios of $F_{S}$ / $F_{L}$ along each S-L line have been calculated to examine the relative 
stability of the S-L lines. In the previous section, the absolute classification efficiencies of the S-L lines for Seyferts 
and LINERs are compared in regions above the Ka03 (Ke01) line to assess the global classification efficiency of different 
S-L lines. However, the comparisons of absolute efficiency cannot reveal the classification efficiency of the S-L lines in 
local regions and their stability along the S-L lines. To further investigate the classification efficiency of the S-L lines 
in various local regions and their stability, the S-L lines are divided into several local regions based on different ways, 
and the $F_{S}$ and $F_{L}$ values for these regions are calculated separately. By analyzing the variation of the ratios of 
$F_{S}$ / $F_{L}$ in these local regions, the stability of the classification efficiency of each S-L line across the overall 
range is further compared. Through the previous comparison of the absolute efficiency using $F_{S}$ and $F_{L}$, the SO line 
clearly outperforms the S line and O line. The S line and O line, while useful for classification, are excluded from this 
comparison due to their lower classification efficiency than the SO line.

	The 5,791 Seyferts and 4,228 LINERs classified in both the [S~{\sc ii}] and the [O~{\sc i}] BPT diagram are represented in 
the [N~{\sc ii}] BPT diagram. As shown in panel (a) of Fig.~\ref{fig:fluctuation}, the long-bins defined in Section \ref{result} 
divided the contour maps for Seyferts and LINERs into 27 groups along the S-L lines. Meanwhile, as shown in panels (b) and (c) 
of Fig.~\ref{fig:fluctuation}, 25 arcs in cyan are translated 0.04 dex equally by Ke01 line and Ka03 line along the direction 
of log(\oiii)=log(\nii), dividing the contour maps for Seyferts and LINERs into 24 Ke-long-bins and Ka-long-bins along the S-L 
lines, respectively. Then the values of $F_{S}$ / $F_{L}$ within each long-bin, Ke-long-bin and Ka-long-bin are calculated. The 
dependencies of $F_{S}$ / $F_{L}$ on the long-bins, Ke-long-bins and Ka-long-bins are shown in panels (d), (e) and (f) of 
Fig.~\ref{fig:fluctuation}, respectively.

	It is obvious that the $F_{S}$ / $F_{L}$ for the SO line, the Sc07 line and the Fe10 line show no obvious difference 
in the region far above the Ke01 line (log(\nii) larger than 0). However, regardless of long-bins, Ke-long-bins, or Ka-long-bins, 
the $F_{S}$ / $F_{L}$ for the SO line in the composite galaxy region, which is above the Ka03 line and below the Ke01 line, is 
much closer to 1 than for those for the Sc07 line and the Fe10 line, indicating the consistency of the efficiency of the SO line 
in classifying Seyferts and LINERs in the composite galaxy region and the region above the Ke01 line.

\subsection{Comparisons of confidence bands}

\begin{figure*}
    \centering
	 \includegraphics[width=2\columnwidth]{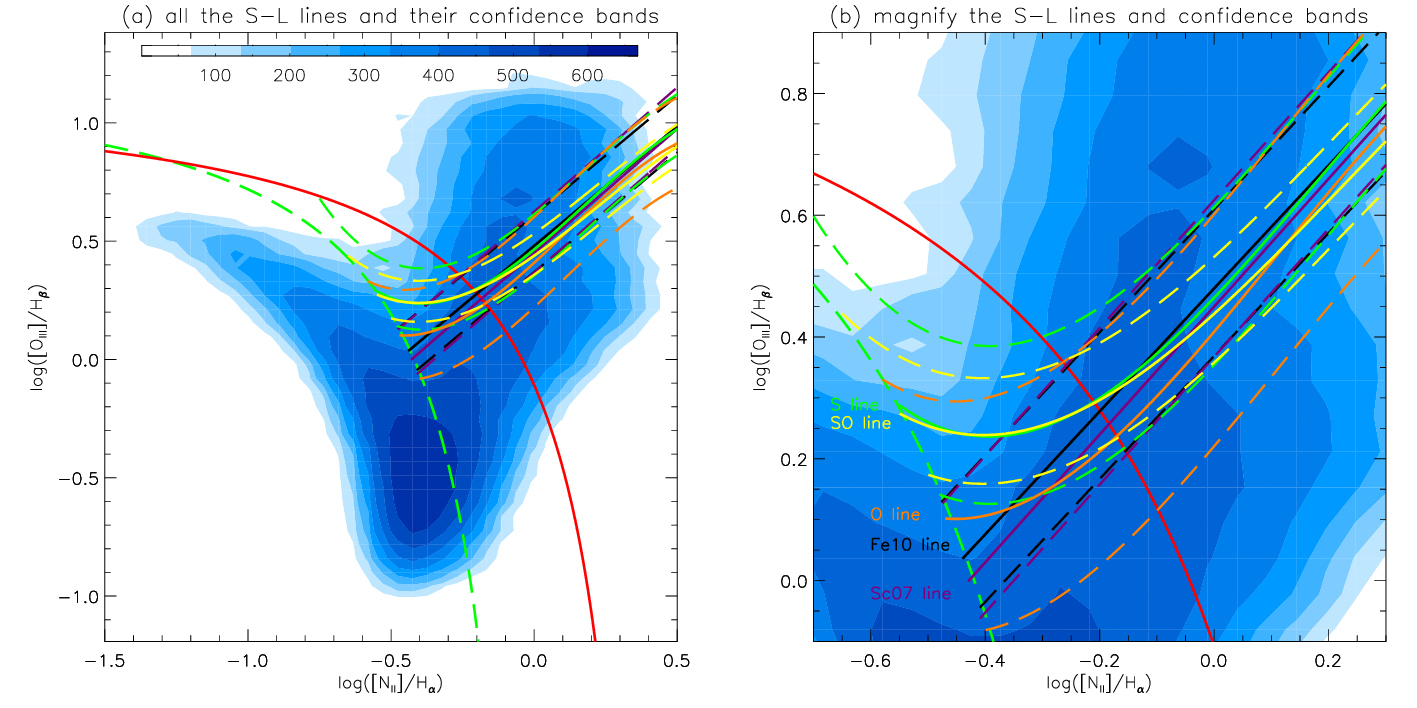}
	 \caption{This figure illustrates the comparison of all the versions of the S-L lines and their 
    confidence bands mentioned in this manuscript, in the [N~{\sc ii}] BPT diagram. (a) shows the S line
    (solid green line), the O line (solid orange line), the SO line (solid yellow line), the Sc07 line 
    (solid purple line) and the Fe10 line (solid black line), and the corresponding confidence bands 
    in dashed lines in the same color. The Ke01 line and the Ka03 line have the same line styles as in 
    Fig.~\ref{fig:s-bin}. The contour filled with bluish colors represents the results for the sample of 
    our collected galaxies. The corresponding number densities of the different colors are shown in the 
    colorbar on the top of the panel. (b) shows the S-L lines in the [N~{\sc ii}] BPT diagram within 
    smaller limited ranges of log(\nii) and log(\oiii).}
	 \label{fig:alltogether}
\end{figure*}

	For the third, besides the effectiveness of the S-L lines checked by $F_{S}$ and $F_{L}$ and their stability checked by 
$F_{S}$ / $F_{L}$, confidence bands have been determined for S-L lines. The upper boundary of the confidence bands ensures that 
LINERs classified in the [S~{\sc ii}] and [O~{\sc i}] BPT diagram above the Ka03 line and above this line account for not larger 
than 2$\%$ of the total LINERs classified in the [S~{\sc ii}] and [O~{\sc i}] BPT diagram above the Ka03 line. Meanwhile, the 
lower boundary of the confidence bands ensures that Seyferts classified in the [S~{\sc ii}] and [O~{\sc i}] BPT diagram above 
the Ka03 line and below this line account for not larger than 2$\%$ of the total Seyferts classified in the [S~{\sc ii}] and 
[O~{\sc i}] BPT diagram above the Ka03 line, as shown in panels (d) of Fig.~\ref{fig:s-bin}, Fig.~\ref{fig:o-bin}, Fig.~\ref{fig:so-bin}. 
The initial threshold is set at 99.73$\%$ (3$\sigma$), however, several points in the remaining 0.27$\%$ that are far from the 
main distribution of the [N~{\sc ii}] BPT diagram can significantly interfere with the positions of confidence bands, making the 
comparisons less meaningful. Meanwhile, the S-L lines can accurately classify approximately 96$\%$ of Seyferts and LINERs as the 
largest $F_{S}$ and $F_{L}$ are around 94$\%$, therefore, the threshold should be set above 96$\%$. Consequently, the threshold 
is adjusted to 98$\%$ (2.33$\sigma$) to ensure the confidence bands are reliable and comparable.

	The upper and lower boundaries of the confidence bands for the S line are obtained by the S line plus 0.1486 and minus 
0.1105, respectively. For the O line, the upper and lower boundaries of the confidence bands are obtained by the O line plus 0.1930 
and minus 0.1880, respectively. The upper and lower boundaries of the confidence bands for the Sc07 line are obtained by the 
Sc07 line plus 0.1748 and minus 0.0824, respectively. For the Fe10 line, the upper and lower boundaries of the confidence bands 
are obtained by the Fe10 line plus 0.1300 and minus 0.1110, respectively. Impressively, for the SO line, the upper and lower 
boundaries of the confidence bands are obtained by the SO line plus 0.0935 and minus 0.0800 merely, respectively. The upward, 
downward and total (upward $+$ downward) shift distances of the S-L lines are presented in Table~\ref{tab:errbar}. 

\begin{table}[ht!]
	 \centering
    \caption{\centering Upward, Downward and Total Shift Distances of S-L Lines for Obtaining Confidence Bands}
	 \begin{tabular}{lccc}
		\hline
      \noalign{\smallskip} 
                & Upward shift & Downward shift & Total shift \\
		\noalign{\smallskip}
      \hline
      \noalign{\smallskip}
       S line    & 0.1486        & 0.1105        & 0.2591      \\
       O line    & 0.1930        & 0.1880        & 0.3810      \\
       SO line   & 0.0935        & 0.0800        & 0.1735      \\
       Sc07 line & 0.1748        & 0.0824        & 0.2572      \\
       Fe10 line & 0.1300        & 0.1110        & 0.2410      \\
		\noalign{\smallskip}
      \hline
      \noalign{\smallskip}
	 \end{tabular}
    \tablefoot{This table demonstrates the values for upward, downward and total shift distances of each S-L line.
         Column 1 shows the different S-L lines. 
	      Column 2 shows the upward shift distance of the upper boundary of the confidence bands for each S-L line.
	      Column 3 shows the downward shift distance of the lower boundary of the confidence bands for each S-L line. 
	      Column 4 shows the total shift distance of the upper and lower boundaries of the confidence bands for each S-L line.}
\label{tab:errbar}
\end{table}
The smallest total shift distance of the upper and lower boundaries of the confidence bands for the SO line makes the SO line 
far superior to the other S-L lines, as shown in Fig.~\ref{fig:alltogether}.

\subsection{Optimal choice}

	Considering the above aspects, the SO line is preferred in the [N~{\sc ii}] BPT diagram. Moreover, as described in Section 
\ref{sql}, we have demonstrated through SQL search queries that the number of galaxies with reliable [N~{\sc ii}] emission lines 
is much higher than those of galaxies with reliable [S~{\sc ii}] or [O~{\sc i}] emission lines. This indicates that the calibrated 
SO line in the [N~{\sc ii}] BPT diagram can help classify a large number of galaxies more precisely and efficiently into Seyferts 
and LINERs in the [N~{\sc ii}] BPT diagram.

\begin{figure}[ht!]
\centering
\includegraphics[width=1\columnwidth]{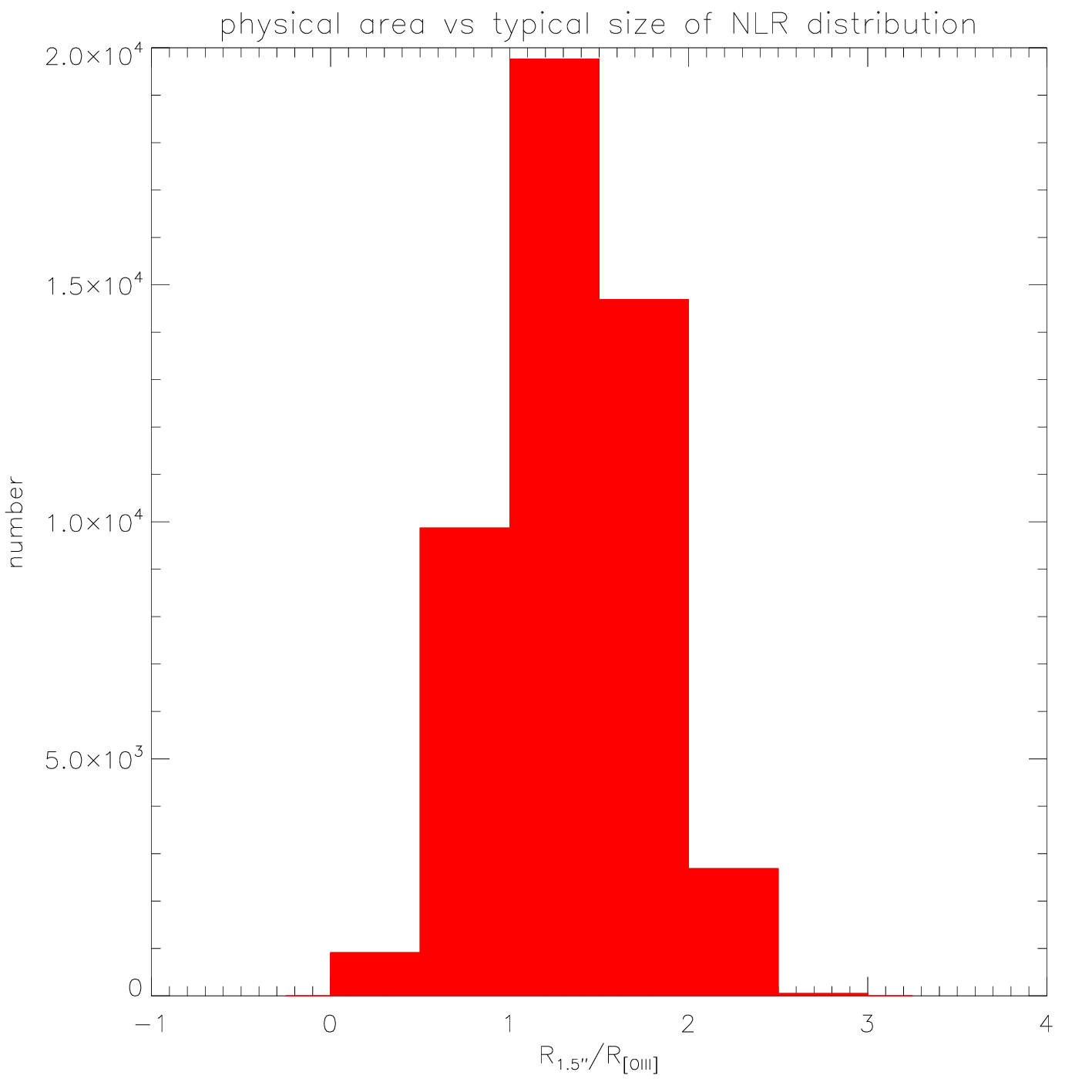}
\caption{This figure shows the distribution of the ratio $R_{1.5^{\prime\prime}} / R_{\text{[O~{\sc iii}]}}$ for objects in Sample 1.}
\label{fig:aperture}
\end{figure}

It is worth noting that \cite{2014MNRAS.441.2296M} have pointed out that the positions of a galaxy in the BPT diagrams can vary with 
the aperture used for observation, since at low redshifts ($z < 0.45$), only the inner zones of galaxies can be captured, rather than 
their full extents. To examine the impacts of aperture effects on our results, which is collected from SDSS DR16 spectra using a fixed 
$1.5^{\prime\prime}$-radius fiber aperture, the physical area captured by a $1.5^{\prime\prime}$-radius aperture is compared with the 
typical size of the narrow line region (NLR).

First, the physical radius corresponding to a $1.5^{\prime\prime}$-radius aperture ($R_{1.5^{\prime\prime}}$) can be calculated with 
the redshift (z). Second, \citet{2013MNRAS.430.2327L} have provided a correlation between the size of the NLR and [O~{\sc iii}] line 
luminosity ($L_{\text{[O~{\sc iii}]}}$, expressed in $10^{42}~\mathrm{erg~s}^{-1}$):
\begin{equation}
	\log(R_{\text{[O~{\sc iii}]}}) = 0.250 \log(L_{\text{[O~{\sc iii}]}}) + 3.746
\end{equation}, where $R_{\text{[O~{\sc iii}]}}$ is the radius of the NLR of [O~{\sc iii}] emission line, expressed in pc.

The distribution of the ratio $R_{1.5^{\prime\prime}} / R_{\text{[O~{\sc iii}]}}$ for each object is subsequently calculated and shown in 
Fig.~\ref{fig:aperture}. Among our sample 1, the ratios of $R_{1.5^{\prime\prime}}$ to $R_{\text{[O~{\sc iii}]}}$ for 37,188 (78$\%$) objects 
are greater than 1, for which the NLRs are fully captured by the $1.5^{\prime\prime}$ aperture. Nevertheless, there should be caution when 
applying the SO line to datasets with significantly different apertures.

\subsection{The application of SVM technique}  \label{sec:ml}

\begin{figure}
\centering
\includegraphics[width=1.0\columnwidth]{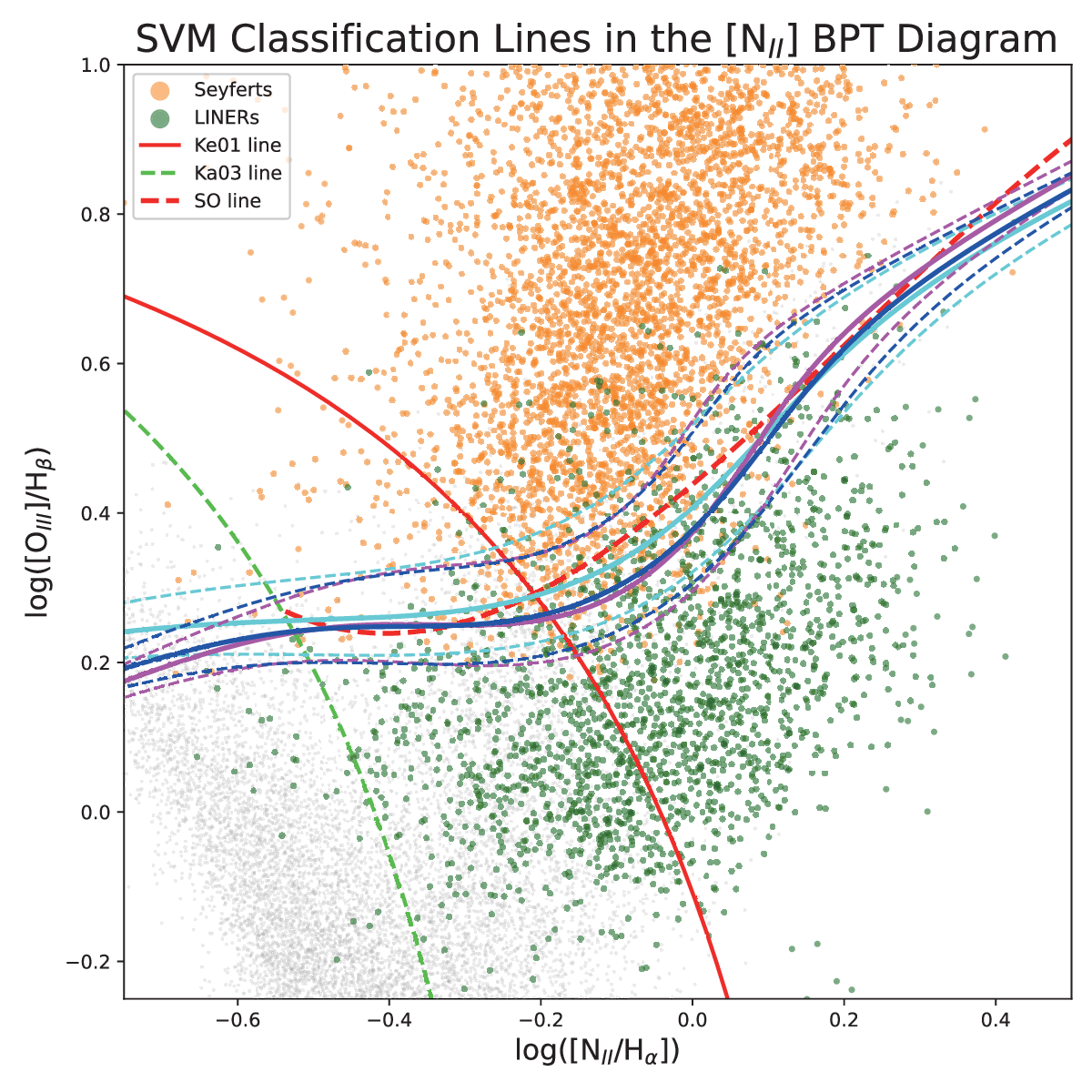}
\caption{This figure shows the three SVM classification lines for Seyferts and LINERs in the [N~{\sc ii}] BPT diagram. 
The solid orange and dark green circles represent the Seyferts and LINERs consistently classified in both the [S~{\sc ii}] BPT diagram and 
the [O~{\sc i}] BPT diagram using Sample 1. The solid red line is the Ke01 line. The dashed green line is the Ka03 line. The dashed red line 
is the SO line. The solid blue, cyan and magenta lines represent the SVM classification lines obtained using Samples 1, 3 and 4, respectively. 
The dashed lines of the corresponding color represent the upper and lower boundaries of margins.}
\label{fig:svm}
\end{figure}

The SVM (Support Vector Machine) is a class of supervised learning methods widely used for classifications, especially binary classifications. 
The SVM supports various kernel functions that can non-linearly map input vectors into a high-dimensional feature space, where non-linear 
problems become linearly separable \citep{10.1162/089976698300017467,708428}. In the feature space, the SVM determines the coefficients of 
the linear hyperplane through the optimization of the Lagrangian multiplier, thereby constructing an optimal hyperplane (decision boundary) 
that maximizes the margin between support vectors \citep{Cortes1995}. 
More detailed introductions to SVM are available on the scikit-learn SVM module page\footnote{\url{https://scikit-learn.org/stable/modules/svm.html\#}}.

Applying the SVM to classify Seyferts and LINERs in the [N~{\sc ii}] BPT diagram offers an independent test of the validity of the SO line. 
Specifically, based on Sample 1, objects classified as Seyferts or LINERs consistently in both the [S~{\sc ii}] and [O~{\sc i}] BPT 
diagrams are selected, and their [O~{\sc iii}]/H$\beta$ and [N~{\sc ii}]/H$\alpha$ ratios are collected and labeled as 1 (Seyferts) and 
0 (LINERs), respectively. These two sets of data are then standardized and input into the SVM model. The kernel is set to \texttt{poly} with 
$degree = 3$, and the penalty parameter is fixed at the default value $C = 1$. The resulting decision boundary is the SVM classification line 
between Seyferts and LINERs in the [N~{\sc ii}] BPT diagram, as shown by the solid blue curve in Fig.~\ref{fig:svm}. Objects located within 
the margin (between the dashed blue curves) are considered to have low classification confidence. The same procedures are then applied to 
Samples 3 and 4, resulting in two additional SVM classification lines. Above the Ka03 line, the positions and shapes of the three SVM 
classification lines are all in good agreement with the SO line, providing strong support for the SO line and further confirming that the 
selection effects have few impacts on determining the SO line.

\subsection{Basic applications of SO line}

\begin{figure*}
	\centering
	\includegraphics[width=2\columnwidth]{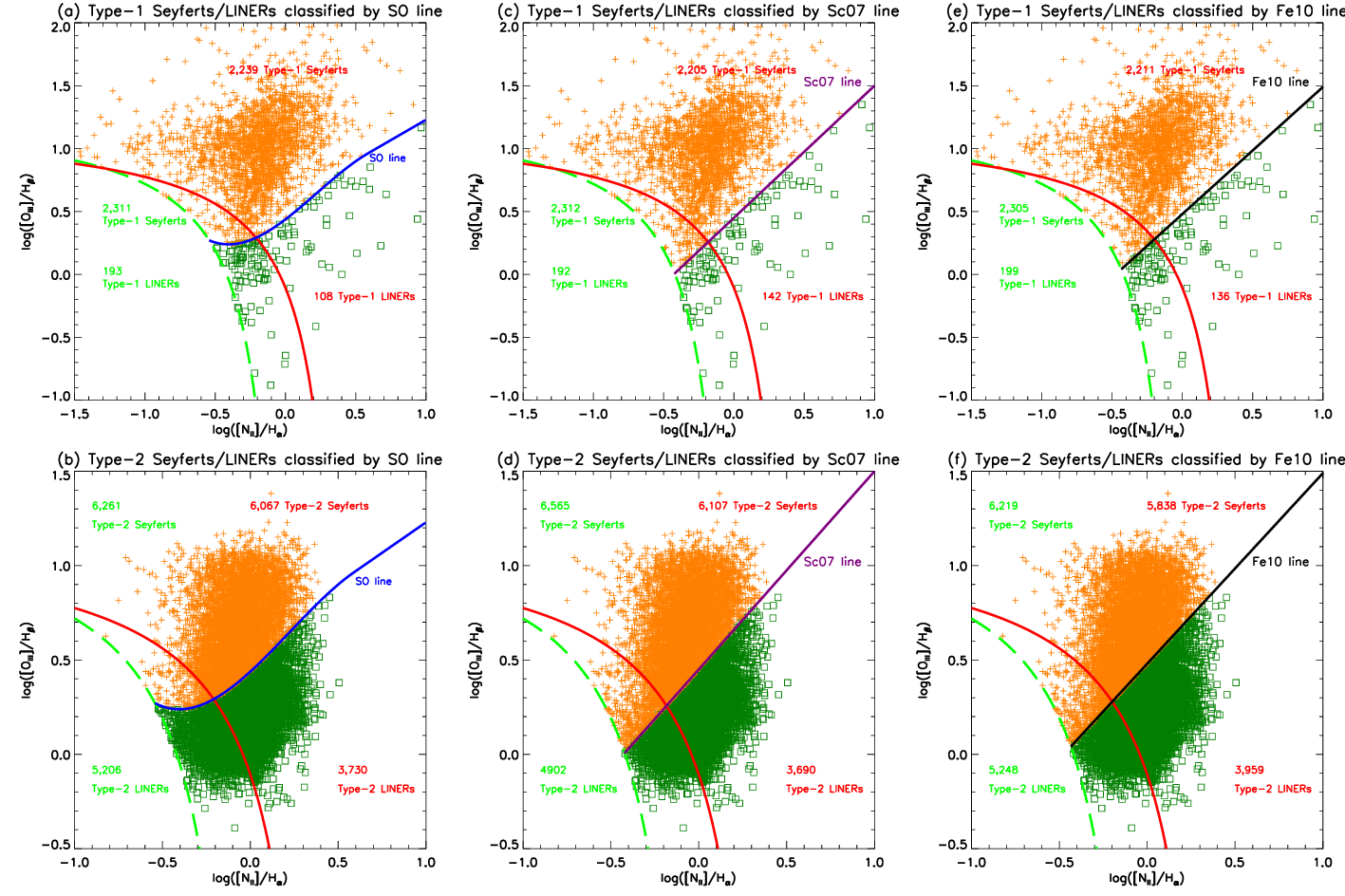}
	\caption{This figure demonstrates the distributions and number counts of Type-1 Seyferts and Type-1 LINERs, as well as Type-2 
Seyferts and Type-2 LINERs, classified by the SO line, Sc07 line and the Fe10 line, respectively, combined with the Ke01 (Ka03) 
line, in the [N~{\sc ii}] BPT diagram. 
(a) shows the distributions of Type-1 Seyferts and Type-1 LINERs classified by the SO line, combined with the Ke01 (Ka03) line, 
in the [N~{\sc ii}] BPT diagram. The samples of Type-1 Seyferts (plus signs in orange) and Type-1 LINERs (hollow squares in dark 
green) are collected from the quasars catalog proposed by \citet{2011ApJS..194...45S}. The solid red line is the Ke01 line. The 
dashed green line is the Ka03 line. The solid blue line is the SO line. The text highlighted in red (green) represents the number 
counts of Type-1 Seyferts and Type-1 LINERs classified by the SO line combined with the Ke01 (Ka03) line. 
(b) shows the distributions of Type-2 Seyferts and Type-2 LINERs classified by the SO line, combined with the Ke01 (Ka03) line, 
in the [N~{\sc ii}] BPT diagram, while also considering the classification results of the Ke01 lines in the [S~{\sc ii}] and 
[O~{\sc i}] BPT diagrams. The samples of Type-2 Seyferts (plus signs in orange) and Type-2 LINERs (hollow squares in dark green) 
are collected in Section \ref{sql}. The other symbols and line styles have the same meanings as those in panel (a). 
(c) shows the distributions of Type-1 Seyferts and Type-1 LINERs classified by the Sc07 line, combined with the Ke01 (Ka03) line, 
in the [N~{\sc ii}] BPT diagram. The solid purple line is the Sc07 line. The other symbols and line styles have the same meanings 
as those in panel (a). 
(d) shows the distributions of Type-2 Seyferts and Type-2 LINERs classified by the Sc07 line, combined with the Ke01 (Ka03) line, 
in the [N~{\sc ii}] BPT diagram, while also considering the classification results of the Ke01 lines in the [S~{\sc ii}] and 
[O~{\sc i}] BPT diagrams. The solid purple line is the Sc07 line. The other symbols and line styles have the same meanings 
as those in panel (b). 
(e) shows the distributions of Type-1 Seyferts and Type-1 LINERs classified by the Fe10 line, combined with the Ke01 (Ka03) line, 
in the [N~{\sc ii}] BPT diagram. The solid black line is the Fe10 line. The other symbols and line styles have the same meanings 
as those in panel (a). 
(f) shows the distributions of Type-2 Seyferts and Type-2 LINERs classified by the Fe10 line, combined with the Ke01 (Ka03) line, 
in the [N~{\sc ii}] BPT diagram, while also considering the classification results of the Ke01 lines in the [S~{\sc ii}] and 
[O~{\sc i}] BPT diagrams. The solid black line is the Fe10 line. The other symbols and line styles have the same meanings 
as those in panel (b).}
	\label{fig:type_ratio}
\end{figure*}

\begin{table*}
    \centering
    \caption{\centering Number Counts and Ratios of Seyferts and LINERs classified by S-L Lines in our Basic Applications}
    \begin{tabular}{lccccccccccc}
        \hline
        \noalign{\smallskip}
        S-L line   & $N_{S1}$ & $N_{S2}$ & $N_{L1}$ & $N_{L2}$ & $N_{S1} / N_{L1}$ & $N_{S2} / N_{L2}$  & $P_{L_{ex}}$ ($\%$) & $\frac{N_{L1} \times N_{S2}}{N_{S1}}$ & $n_{1}$ & $n_{2}$\\
        \noalign{\smallskip}
        \hline
        \noalign{\smallskip}
        \multicolumn{11}{c}{Ka03} \\
        \noalign{\smallskip}
        SO line    & 2,311    & 6,261    & 193      & 5,206    & 11.97             & 1.20               & 89.96 & 523 & 1,263 & 4,322           \\
        Sc07 line  & 2,312    & 6,565    & 192      & 4,902    & 12.04             & 1.34               & 88.88 & 545 & 1,135 & 4,023           \\
        Fe10 line  & 2,305    & 6,219    & 199      & 5,248    & 11.58             & 1.19               & 89.77 & 537 & 1,274 & 4,337           \\
        \noalign{\smallskip}
        \hline
        \noalign{\smallskip}
        \multicolumn{11}{c}{Ke01} \\
        \noalign{\smallskip}
        SO line    & 2,239    & 6,067    & 108      & 3,730    & 20.73             & 1.63               & 92.15 & 293 & 927  & 3,279           \\
        Sc07 line  & 2,205    & 6,107    & 142      & 3,690    & 15.53             & 1.66               & 89.34 & 393 & 875  & 3,097           \\
        Fe10 line  & 2,211    & 5,838    & 136      & 3,959    & 16.26             & 1.47               & 90.93 & 359 & 989  & 3,391           \\
        \noalign{\smallskip}
        \hline
        \noalign{\smallskip}
    \end{tabular}
    \tablefoot{This table demonstrates the Seyferts and LINERs classification results in Fig.~\ref{fig:type_ratio}. 
The upper/lower part of the table describes the classification results for galaxies lying above the Ka03/Ke01 
line in the [N~{\sc ii}] BPT diagram. Column 1 shows the different S-L lines. Column 2 shows the 
$N_{S1}$. Column 3 shows the $N_{S2}$. Column 4 shows the $N_{L1}$. Column 5 shows the $N_{L2}$. Column 6 shows 
the $N_{S1} / N_{L1}$. Column 7 shows the $N_{S2} / N_{L2}$. Column 8 shows the $P_{L_{ex}}$, with the results in percentage form. 
Column 9 shows the $\frac{N_{L1} \times N_{S2}}{N_{S1}}$. Column 10 shows the $n_{1}$. Column 11 shows the $n_{2}$.}
    \label{tab:nonAGN-LINERs}
\end{table*}

     The proposed and preferred SO line in the [N~{\sc ii}] BPT diagram is then applied to the following project.
 
     Similar to the criteria in Section \ref{sql} (redshift z $<$ 0.35, narrow emission line fluxes of [O~{\sc iii}]$\lambda$5007\AA, 
H$\beta$, [N~{\sc ii}]$\lambda$6586\AA\ and H$\alpha$ being 5 times larger than their corresponding uncertainties which are greater 
than 0, and the median S/N per pixel for the restframe 6400-6765\AA~ and 4750-4950\AA~ regions, which are near the H$\alpha$ and 
H$\beta$ emission lines, being greater than 10), there are 2,703 quasars (Type-1 AGNs) being collected from the quasar catalog 
reported by \citet{2011ApJS..194...45S}, which is distributed as a FITS file. Our analysis specifically utilizes the following 
fields in the FITS file:
\begin{lstlisting}
REDSHIFT, 
LOGL_OIII_5007, LOGL_OIII_5007_ERR, 
LOGL_NARROW_HB, LOGL_NARROW_HB_ERR, 
LOGL_NII_6585,  LOGL_NII_6585_ERR, 
LOGL_NARROW_HA, LOGL_NARROW_HA_ERR, 
LINE_MED_SN_HA, LINE_MED_SN_HB.
\end{lstlisting}

Then, the Type-1 AGNs collected from \citet{2011ApJS..194...45S} and all the Type-2 objects in sample 1 are shown in the 
[N~{\sc ii}] BPT diagram, respectively. Those objects located above the Ka03 (Ke01) line and above the SO line in the [N~{\sc ii}] BPT 
diagram are classified as Type-1 and Type-2 Seyferts, respectively, while those located above the Ka03 (Ke01) line but below the SO 
line are classified as Type-1 and Type-2 LINERs.
The number counts of Type-1 Seyferts ($N_{S1}$ = 2,311 (2,239)), Type-2 Seyferts ($N_{S2}$ = 6,261 (6,067)), Type-1 LINERs ($N_{L1}$ = 
193 (108)) and Type-2 LINERs ($N_{L2}$ = 5,206 (3,730)) have been calculated. The numbers out of parentheses correspond to the number 
counts obtained based on the application of the SO line combined with the Ka03 line, while those in the parentheses correspond to the 
number counts obtained based on the application of the SO line combined with the Ke01 line. These results are highlighted in green and 
red respectively in panel (a) and (b) of Fig.~\ref{fig:type_ratio}, and are also listed in the third and seventh rows of 
Table~\ref{tab:nonAGN-LINERs}. It is worth mentioning that in the composite galaxy region (above the Ke01 line and below the Ka03 line 
in the [N~{\sc ii}] BPT diagram), only objects that are lying above the Ke01 lines in the [S~{\sc ii}] and [O~{\sc i}] BPT diagrams are 
selected. Without these constraints, the numbers of Type-2 LINERs increase by approximately 10,000 objects, while those of Type-2 
Seyferts increase by only around 1,000, as shown in Section \ref{result}. This selection ensures more reliable Type-2 LINER samples in 
the composite galaxy region by reducing objects that fail to meet AGN criteria using three BPT diagrams.

    As our preliminary results, based on the application of the SO line combined with the Ka03 (Ke01) line, the number ratio of Type-1 
Seyferts to Type-1 LINERs ($N_{S1} / N_{L1}$ = 11.97 (20.73)) is 9.98 (12.72) times greater than that of Type-2 Seyferts to Type-2 LINERs 
($N_{S2} / N_{L2}$ = 1.20 (1.63)). As shown in the third and seventh rows of Table~\ref{tab:nonAGN-LINERs}, these results provide further 
clues to support that about 89.96$\%$ (92.15$\%$) of objects should be excluded from the Type-2 LINER sample or reclassified. Meanwhile, 
the number ratios of Type-1 Seyferts to Type-1 LINERs and Type-2 Seyferts to Type-2 LINERs based on the Sc07 line and Fe10 line have also 
been calculated. Based on the application of the Sc07 line combined with the Ka03 (Ke01) line, the calculated $N_{S1} / N_{L1}$ = 12.04 
(15.53) is 8.99 (9.36) times larger than $N_{S2} / N_{L2}$ = 1.34 (1.66), while based on the application of the Fe10 line combined with 
the Ka03 (Ke01) line, the calculated $N_{S1} / N_{L1}$ = 11.58 (16.26) is 9.73 (11.06) times larger than $N_{S2} / N_{L2}$ = 1.19 (1.47). 
As shown in panels (c), (d), (e) and (f) of Fig.~\ref{fig:type_ratio} and the fourth, fifth, eighth and ninth rows of 
Table~\ref{tab:nonAGN-LINERs}, these results indicate that the same proportion of objects should be excluded from the Type-2 LINER sample 
or reclassified compared to the SO line.

    For the differences in the two number ratios, at least three points can be made to explain this. 
\begin{enumerate}

    \item First, as mentioned in Section \ref{introduction}, a significant fraction of objects in the Type-2 LINER sample are indeed 
non-AGN-related. Although these non-AGN-related LINERs lie above the Ke01 (Ka03) line and below the SO line in the [N~{\sc ii}] BPT diagram, 
they should not be included in AGN statistics. 
Therefore, when calculating the number ratio of Type-2 Seyferts to Type-2 LINERs, they should 
be excluded. Under the assumption that part of Type-2 LINERs are non-AGN-related, considering that some objects should be excluded from the 
Type-2 LINER sample could be naturally applied to explain the number ratio of Type-2 Seyferts to Type-2 LINERs, which is quite different from 
that of Type-1 Seyferts to Type-1 LINERs, as follows. 

In calculating the proportions of objects that should be excluded from the Type-2 LINER samples or reclassified ($P_{L_{ex}}$), the following 
formula is used:
\begin{equation}
		P_{L_{ex}} = \frac{N_{L2} - \frac{N_{L1} \times N_{S2}}{N_{S1}}}{N_{L2}}
\end{equation}, and the resulting values are presented in column 8 of Table~\ref{tab:nonAGN-LINERs}. By this formula, the percentage of the 
difference between the actual count ($N_{L2}$) and the theoretical count ($\frac{N_{L1} \times N_{S2}}{N_{S1}}$, based on the assumption that 
$N_{S1} / N_{L1}$ = $N_{S2} / N_{L2}$, and the resulting values are presented in column 9 of Table~\ref{tab:nonAGN-LINERs}) of Type-2 LINERs 
relative to the actual count is quantified.

    Based on the application of the defined SO line combined with the Ka03 (Ke01) line, if there are 89.96$\%$ (92.15$\%$) Type-2 LINERs 
that are non-AGN-related, the number ratio of Type-2 Seyferts to Type-2 LINERs should be about 6,261 : 523 = 11.97 (6,067 : 293 = 20.71), 
leading to the same number ratio of Type-1 Seyferts to Type-1 LINERs being 2,311 : 193 = 11.97 (2,239 : 108 = 20.73). Similarly, based on 
the application of the proposed Sc07 line combined with the Ka03 (Ke01) line, if there are 88.88$\%$ (89.34$\%$) Type-2 LINERs that are 
non-AGN-related, the number ratio of Type-2 Seyferts to Type-2 LINERs should be about 6,565 : 545 = 12.05 (6,107 : 393 = 15.54), leading 
to the same number ratio of Type-1 Seyferts to Type-1 LINERs being 2,312 : 192 = 12.04 (2,205 : 142 = 15.53). Based on the application of 
the proposed Fe10 line combined with the Ka03 (Ke01) line, if there are 89.77$\%$ (90.93$\%$) Type-2 LINERs that are non-AGN-related, the 
number ratio of Type-2 Seyferts to Type-2 LINERs should be about 6,219 : 537 = 11.58 (5,838 : 359 = 16.26), leading to the same number 
ratio of Type-1 Seyferts to Type-1 LINERs being 2,305 : 199 = 11.58 (2,211 : 136 = 16.26). The data in this paragraph are shown in 
Table~\ref{tab:nonAGN-LINERs}.

\cite{2025A&A...693A..95D} have demonstrated that 72$\%$ of LINERs are ionized predominantly by old stellar populations rather 
than AGN activity, through random forest machine learning combined with the $D4000$ continuum break index and H$\alpha$, 
[O~{\sc iii}]$\lambda$5007\AA~and [N~{\sc ii}]$\lambda$6584\AA~ emission-line features, which is broadly consistent with the above results.

    \item Second, part of objects classified as Type-2 LINERs in the [N~{\sc ii}] BPT diagram may actually be true Type-2 AGNs, as several 
candidates reported by \cite{2015AJ....149...75L, 2010ApJ...714..115S, 2021ApJ...922..248X} and two samples of candidates reported by 
\cite{2014MNRAS.438..557Z, 2016A&A...594A..72P}. In the framework of the unified model of AGNs, the true Type-2 AGNs have the same orientation 
as Type-1 AGNs, with unobscured central regions despite the absence of broad emission lines. Under the assumption that all the 
objects in the existing Type-2 LINER sample are AGN-related, since the non-AGN-related LINERs mentioned earlier cannot currently be precisely 
dismissed, considering part of the objects in the Type-2 LINER sample as Type-1 LINER candidates could be naturally applied to explain the 
number ratio of Type-2 Seyferts to Type-2 LINERs, which is quite different from that of Type-1 Seyferts to Type-1 LINERs, as follows. 

If the numbers of objects expected to be reclassified from the Type-2 LINER samples to the Type-1 LINER samples are $n_{1}$, then the equation
\begin{equation}
	    	\frac{N_{S1}}{N_{L1} + n_{1}} = \frac{N_{S2}}{N_{L2} - n_{1}}
\end{equation} should hold, with the resulting values presented in column 10 of Table~\ref{tab:nonAGN-LINERs}. 

    Based on the application of the defined SO line combined with the Ka03 (Ke01) line, if there are 1,263 (927) Type-2 LINERs that can be 
reclassified as Type-1 LINERs, the number ratio of Type-2 Seyferts to Type-2 LINERs should be about 6,261 : 3,943 = 1.59 (6,067 : 2,803 = 2.16), 
leading to the same number ratio of Type-1 Seyferts to Type-1 LINERs being 2,311 : 1,456 = 1.59 (2,239 : 1,035 = 2.16). Similarly, based on the 
application of the proposed Sc07 line combined with the Ka03 (Ke01) line, if there are 1,135 (875) Type-2 LINERs that can be reclassified as 
Type-2 Seyferts, the number ratio of Type-2 Seyferts to Type-2 LINERs should be about 6,565 : 3,767 = 1.74 (6,107 : 2,815 = 2.17), leading to 
the same number ratio of Type-1 Seyferts to Type-1 LINERs being 2,312 : 1,327 = 1.74 (2,205 : 1,017 = 2.17). Based on the application of the 
proposed Fe10 line combined with the Ka03 (Ke01) line, if there are 1,274 (989) Type-2 LINERs that can be reclassified as Type-2 Seyferts, the 
number ratio of Type-2 Seyferts to Type-2 LINERs should be about 6,219 : 3,974 = 1.56 (5,838 : 2,970 = 1.97), leading to the same number ratio 
of Type-1 Seyferts to Type-1 LINERs being 2,305 : 1,473 = 1.56 (2,211 : 1,125 = 1.97).

    \item Third, according to \cite{2024MNRAS.529.4500M}, based on the Mapping Nearby Galaxies at Apache Point Observatory survey, a fading AGN 
exhibits LINER-like ionization in its central region while showing Seyfert-like emission in the outskirts. This suggests that some AGNs might 
lie above the Ke01 line and below the SO line in the [N~{\sc ii}] BPT diagram and be classified as LINERs, with their Seyfert-like nature 
potentially masked by the central LINER-like properties. Consequently, part of objects in the Type-2 LINER sample might need to be reclassified 
into the Type-2 Seyfert sample. Similarly, under the assumption that all the objects in the existing Type-2 LINER sample are AGN-related, 
considering that part of the objects in the Type-2 LINER sample as Type-2 Seyfert candidates could be naturally applied to explain the number ratio 
of Type-2 Seyferts to Type-2 LINERs, which is quite different from that of Type-1 Seyferts to Type-1 LINERs, as follows. 

If the numbers of objects expected to be reclassified from the Type-2 LINER samples to the Type-2 Seyfert samples are $n_{2}$, then the equation
\begin{equation}
	    	\frac{N_{S1}}{N_{L1}} = \frac{N_{S2} + n_{2}}{N_{L2} - n_{2}}
\end{equation} should hold, with the resulting values presented in column 11 of Table~\ref{tab:nonAGN-LINERs}. 

    Based on the application of the defined SO line combined with the Ka03 (Ke01) line, if there are 4,322 (3,279) Type-2 LINERs that can be 
reclassified as Type-2 Seyferts, the number ratio of Type-2 Seyferts to Type-2 LINERs should be about 10,583 : 884 = 11.97 (9,346 : 451 = 20.72), 
leading to the same number ratio of Type-1 Seyferts to Type-1 LINERs being 2,311 : 193 = 11.97 (2,239 : 108 = 20.73). Similarly, based on the 
application of the proposed Sc07 line combined with the Ka03 (Ke01) line, if there are 4,023 (3,097) Type-2 LINERs that can be reclassified as 
Type-2 Seyferts, the number ratio of Type-2 Seyferts to Type-2 LINERs should be about 10,588 : 879 = 12.05 (9,204 : 593 = 15.52), leading to 
the same number ratio of Type-1 Seyferts to Type-1 LINERs being 2,312 : 192 = 12.04 (2,205 : 142 = 15.53). Based on the application of the 
proposed Fe10 line combined with the Ka03 (Ke01) line, if there are 4,337 (3,391) Type-2 LINERs that can be reclassified as Type-2 Seyferts, the 
number ratio of Type-2 Seyferts to Type-2 LINERs should be about 10,556 : 911 = 11.59 (9,229 : 568 = 16.25), leading to the same number ratio 
of Type-1 Seyferts to Type-1 LINERs being 2,305 : 199 = 11.58 (2,211 : 136 = 16.26).

\end{enumerate}
At this stage, for the applications of the three methods to approximately 90$\%$ of the objects in the Type-2 LINER sample, aimed at achieving 
consistency between the number ratio of Type-2 Seyferts to Type-2 LINERs and that of Type-1 Seyferts to Type-1 LINERs, it remains unclear which 
of the methods, or a combination of them, is responsible, and the respective contribution of each. Detailed discussions on comparisons between 
Type-2 LINERs and Type-1 LINERs in larger samples can be found in our being prepared manuscript in the near future.

\section{Main Summary and Conclusions} \label{sec:Conclusion}

	Satisfactory results on the S-L lines have been obtained in the [S~{\sc ii}] and [O~{\sc i}] BPT diagrams using our 
high-quality samples, which are consistent with \citet{2006MNRAS.372..961K}. However, distinct double-peaked features cannot be 
obtained, making it infeasible to define the S-L line in the [N~{\sc ii}] BPT diagram. Meanwhile, the Sc07 proposed by 
\citet{2007MNRAS.382.1415S} and the Fe10 proposed by \cite{2010MNRAS.403.1036C} in the [N~{\sc ii}] BPT diagram are all linear 
and different, while a non-linear line is visually better to describe the intersection boundary for Seyferts and LINERs. To 
address this, a new method is proposed, resulting in our improved Seyfert-LINER classification line, the SO line. With the more 
precise SO line, galaxies with weak [S~{\sc ii}] or [O~{\sc i}] narrow forbidden emission lines but apparent [N~{\sc ii}] emission 
lines can be well classified in the [N~{\sc ii}] BPT diagram. This significantly enhances the efficiency of classifying between 
Type-2 Seyferts and Type-2 LINERs, compared to the Sc07 line and the Fe10 line, and leads to cleaner and larger samples of 
Type-2 Seyferts and Type-2 LINERs. Additionally, the representativeness of Sample 1 and the robustness of the new method 
are confirmed by excluding the impacts of selection and aperture effects, achieved respectively by adjusting the lower limits of 
the flux-to-uncertainty ratio or $SNmedian$, and by testing the $R_{1.5^{\prime\prime}}$ / $R_{\text{[O~{\sc iii}]}}$ ratio. 
The reliability of the SO line is further validated using the SVM technique.

In our basic application, through the proposed SO line, cleaner samples of Type-1 
Seyferts/LINERs and Type-2 Seyferts/LINERs have been created, and the number ratio of Type-2 Seyferts to Type-2 LINERs is 
obviously different from that of Type-1 Seyferts to Type-1 LINERs, indicating that about 90$\%$ Type-2 LINERs are non-AGN-related, 
true Type-2 AGNs or objects exhibiting both Seyfert and LINER characteristics. In the near future, cleaner samples of Type-1/2 
LINERs can be used to provide further clues on the probable number ratio of AGN-related LINERs to non-AGN-related LINERs, and 
cleaner samples of Type-1/2 Seyferts can be used to test the unified model of AGNs with as few effects of LINERs as possible.

\begin{acknowledgements}
      We gratefully acknowledge the anonymous referee for giving us constructive comments and suggestions to greatly improve the manuscript.
      Zhang gratefully acknowledges the kind financial support from GuangXi University, and the grant support from NSFC-12173020 
and NSFC-12373014. Cheng \& Chen gratefully acknowledge the kind grant support from Innovation Project of Guangxi Graduate Education 
YCSW2024006. This manuscript has made use of the data from the SDSS projects\footnote{\url{http://www.sdss3.org/}}, managed by the 
Astrophysical Research Consortium for the Participating Institutions of the SDSS-III Collaborations and the $poly\_fit$ code.
\end{acknowledgements}

\bibliographystyle{aa}
\bibliography{cpz}

\appendix

\section{The detailed SQL conditions and query}  \label{SQL conditions}

First, the redshifts (z) are smaller than 0.35 
to ensure the narrow emission lines, especially [O~{\sc iii}]$\lambda$5007\AA, [N~{\sc ii}]$\lambda$6584\AA, 
[S~{\sc ii}]$\lambda$6717,6731\AA, [O~{\sc i}]$\lambda$6300\AA, H$\beta$ and H$\alpha$, are included in the SDSS spectra, due to the 
narrow emission lines being the ones applied in the following BPT diagrams. Second, the spectral signal-to-noise ratios (S/N) are  
greater than 10 to ensure high-quality spectra. Third, the stellar velocity dispersions are greater than 80 km/s and smaller than 
350 km/s, with at least three times larger than their corresponding uncertainties, to ensure apparent absorption features leading to 
reliable measurements of host galaxy contributions, then to support the reliability of the measurements of the narrow emission lines. 
Fourth, the fluxes of the narrow emission lines applied in the BPT diagrams are at least 5 times larger than their corresponding 
uncertainties in order to confirm the reliability of the narrow emission lines. Then, the SQL Search query applied in detail is as 
follows:
\begin{lstlisting}
SELECT S.plate, S.fiberid, S.mjd, S.z, 
S.snmedian, S.veldisp, S.veldisperr,
G.h_alpha_flux, G.h_alpha_flux_err, 
G.h_beta_flux, G.h_beta_flux_err,
G.oiii_5007_flux, G.oiii_5007_flux_err, 
G.oiii_5007_flux / G.h_beta_flux as o3hb,
G.nii_6584_flux, G.nii_6584_flux_err, 
G.nii_6584_flux / G.h_alpha_flux as n2ha,
G.sii_6717_flux, G.sii_6717_flux_err, 
G.sii_6731_flux, G.sii_6731_flux_err,
(G.sii_6731_flux + 
G.sii_6717_flux) / G.h_alpha_flux as s2ha,
G.oi_6300_flux, G.oi_6300_flux_err, 
G.oi_6300_flux / G.h_alpha_flux as o1ha
FROM GalSpecLine as G JOIN SpecObjall as S 
ON S.specobjid = G.specobjid   
WHERE S.class = 'GALAXY' 
and S.SNmedian > 10 and S.z < 0.35 and 
S.zwarning = 0 and S.veldisperr > 0 
and S.veldisp > 3*S.veldisperr and 
(S.veldisp > 80 and S.veldisp < 350) 
and G.h_beta_flux_err > 0 and 
G.h_beta_flux > 5*G.h_beta_flux_err 
and G.h_alpha_flux_err > 0 and 
G.h_alpha_flux > 5*G.h_alpha_flux_err 
and G.oiii_5007_flux_err > 0 and 
G.oiii_5007_flux > 5*G.oiii_5007_flux_err 
and G.nii_6584_flux_err > 0 and 
G.nii_6584_flux > 5*G.nii_6584_flux_err 
and G.oi_6300_flux_err > 0 and 
G.oi_6300_flux > 5*G.oi_6300_flux_err 
and G.sii_6717_flux_err > 0 and 
G.sii_6717_flux > 5*G.sii_6717_flux_err 
and G.sii_6731_flux_err > 0 and 
G.sii_6731_flux > 5*G.sii_6731_flux_err
\end{lstlisting}
In this query, the SDSS provided GalSpecLine database\footnote{\href{https://skyserver.sdss.org/dr16/en/help/browser/browser.aspx?cmd=description+galSpecLine+U\#\&\&history=description+galSpecLine+U}{https://skyserver.sdss.org/dr16/en/help/browser/browser.aspx?cmd=\allowbreak description+galSpecLine+U\#\&\&history=description+galSpecLine+U}}
includes the emission line measurements obtained from the MPA-JHU research group\footnote{\href{https://www.sdss4.org/dr12/spectro/galaxy\_mpajhu/}{https://www.sdss4.org/dr12/spectro/galaxy\_mpajhu/}}, with applications of multi-Gaussian functions to describe 
narrow emission lines after subtraction of host galaxy contributions determined by the Bruzual and Charlot population synthesis models 
\citep{2003MNRAS.344.1000B}. More detailed descriptions can be found in \cite{2003MNRAS.341...33K,2004ApJ...613..898T,2004MNRAS.351.1151B}.
The SDSS provided SpecObjAll database\footnote{\href{https://skyserver.sdss.org/dr16/en/help/browser/browser.aspx?cmd=description+SpecObjAll+U\#\&\&history=description+SpecObjAll+U}{https://skyserver.sdss.org/dr16/en/help/browser/browser.aspx?cmd=\allowbreak description+SpecObjAll+U\#\&\&history=description+SpecObjAll+U}} contains all the spectroscopic information and the measured basic parameters, especially the information of redshift, stellar velocity 
dispersion, S/N, etc.

\section{Reason for proposing new methods} \label{sec:appendix}

	After collecting narrow emission line galaxies using an SQL Search query from the SDSS DR16, 
we firstly attempt to reproduce the S-L lines in the [S~{\sc ii}] and [O~{\sc i}] BPT diagrams as 
shown in Fig.~4 in \citet{2006MNRAS.372..961K} using the method proposed in their paper, in order 
to test whether the S-L lines will change with variations in sample size, and whether the same 
method can be applied to determine the S-L line in the [N~{\sc ii}] BPT diagram.

\begin{figure*}
	\centering
	\includegraphics[width=2.0\columnwidth]{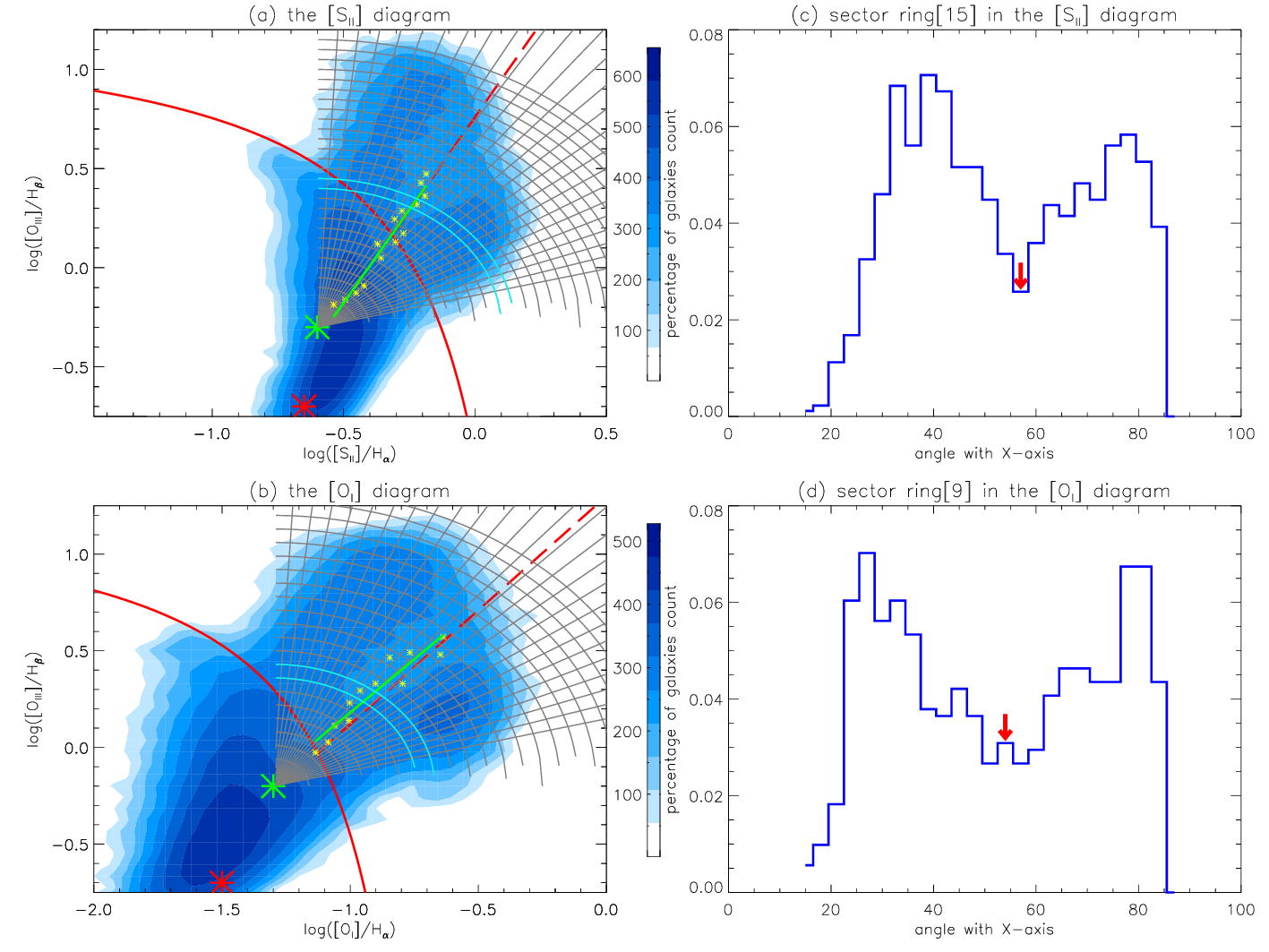}
	\caption{This figure demonstrates the complete process for redefining the S-L lines in the
    [S~{\sc ii}] and [O~{\sc i}] BPT diagrams using the method proposed in \citet{2006MNRAS.372..961K}.
    (a) shows the result of redefining the S-L line in the [S~{\sc ii}] BPT diagram using the same 
    method in \citet{2006MNRAS.372..961K}. The huge red and green asterisks represent the empirical
    base points selected by \citet{2006MNRAS.372..961K} and our new selection. The small yellow 
    asterisks represent the boundary points of the Seyferts and LINERs distributions in each sector 
    ring. The solid red line is the Ke01 line, the dashed red line is the S-L line in the [S~{\sc ii}]
    BPT diagram from \citet{2006MNRAS.372..961K}, and the solid green line is the result of our 
    replication of their work. The contour filled with bluish colors represents the results for 
    the sample of our collected galaxies. The corresponding number densities of the different 
    colors are shown in the colorbar on the right side of the panel. (b) shows the result of 
    redefining the S-L line in the [O~{\sc i}] BPT diagram using the same method in 
    \citet{2006MNRAS.372..961K}. The symbols and line styles have the same meanings as those in 
    panel (a). The contour filled with bluish colors represents the results for the sample of 
    our collected galaxies. The corresponding number densities of the different colors are shown
    in the colorbar on the right side of the panel. (c) shows the dependence of the percentage 
    of galaxies count within each Ke-bin in sector ring[15] relative to the total count of galaxies
    in the entire sector ring on the angle with the X-axis. The red arrow represents the boundary
    point between Seyferts and LINERs within sector ring [15]. (d) shows the dependence of the 
    percentage of galaxies count within each Ke-bin in sector ring[9] relative to the total count
    of galaxies in the entire sector ring on the angle with the X-axis. The red arrow represents
    the boundary point between Seyferts and LINERs within sector ring [9].}
	\label{fig:reke06.eps}
\end{figure*}

	Firstly, two empirical base points are selected manually, labeled as p\textsubscript{[S~{\sc ii}]}
with (log(\sii), log(\oiii)) = (-0.65, -0.7) in the [S~{\sc ii}] BPT diagram and p\textsubscript{[O~{\sc i}]}
with (log(\oi), log(\oiii)) = (-1.5, -0.7) in the [O~{\sc i}] BPT diagram, based on visual inspection of 
Fig.~2 in \cite{2006MNRAS.372..961K}. Secondly, taking p\textsubscript{[S~{\sc ii}]} 
(p\textsubscript{[O~{\sc i}]}) as the center, a sector is drawn. Arcs are then delineated at 0.1 
dex intervals to form incomplete annular regions, referred to as sector rings. Thirdly, each sector
ring is subdivided into Ke-bins by a group of straight lines emanating from  p\textsubscript{[S~{\sc ii}]}
(p\textsubscript{[O~{\sc i}]}) with a specific angle (e.g., 3 degrees per unit) between adjacent lines.
Fourthly, referring to log(\sii) (log(\oi)) as the X-axis and log(\oiii) as the Y-axis, histograms 
are drawn to depict the dependence of the percentage of galaxies count within each Ke-bin relative 
to the total count of galaxies in the corresponding entire sector ring on the angle with the X-axis.
Fifthly, in each histogram, the local minimum point between two peaks is considered as the boundary
point between Seyferts and LINERs in the corresponding sector ring. Finally, by defining the boundary
points from all sector rings and fitting these boundary points with a straight line, the resulting 
boundary lines are the S-L lines in the [S~{\sc ii}] and [O~{\sc i}] BPT diagrams, respectively.

	Actually, during the process, as long as the empirical base points p\textsubscript{[S~{\sc ii}]}
and p\textsubscript{[O~{\sc i}]} are randomly chosen near the distinct valley lines of the contour 
maps and below the Ke01 lines, it will have few effects on the final determined classification lines.
As shown in panel (a) and panel (b) of Fig.~\ref{fig:reke06.eps}, our new selection of empirical base
points are (log(\sii), log(\oiii)) = (-0.6, -0.3) in the [S~{\sc ii}] BPT diagram and (log(\oi), 
log(\oiii)) = (-1.3, -0.2) in the [O~{\sc i}] BPT diagram. In addition, different steps from 0.05 dex
to 0.1 dex during delineating arcs have few effects on the final classification lines in the 
[S~{\sc ii}] and [O~{\sc i}] BPT diagrams. As shown in panel (a) and panel (b) of 
Fig.~\ref{fig:reke06.eps}, the steps between each sector ring are 0.05 dex and 0.07 dex, and the 
straight lines are calibrated every 3 degrees starting from an angle of 15 degrees with the X-axis.
However, if the width of each sector ring is too wide or too narrow, it will result in too many or
too few galaxies in each sector ring, leading to the galaxy number count distributions within sector
rings being extremely unsmooth. The panel (c) and panel (d) in Fig.~\ref{fig:reke06.eps} depict examples
of the dependence of the percentage of galaxies count within each Ke-bin relative to the total number
of galaxies in the entire sector ring, on the angle with the X-axis. As similar as the Fig.~3 in 
\citet{2006MNRAS.372..961K}, the results clearly show double-peaked features. The peak on the side of
the smaller angle with the X-axis represents the densest part of the LINERs, while the peak on the 
side of the larger angle represents the densest part of the Seyferts. The re-determined S-L lines in 
the [S~{\sc ii}] and [O~{\sc i}] BPT diagrams are completely similar to the results in 
\citet{2006MNRAS.372..961K}, strongly indicating that the varying sample size does not significantly 
affect the S-L lines.

\begin{figure*}
	\centering 
	\includegraphics[width=2.0\columnwidth]{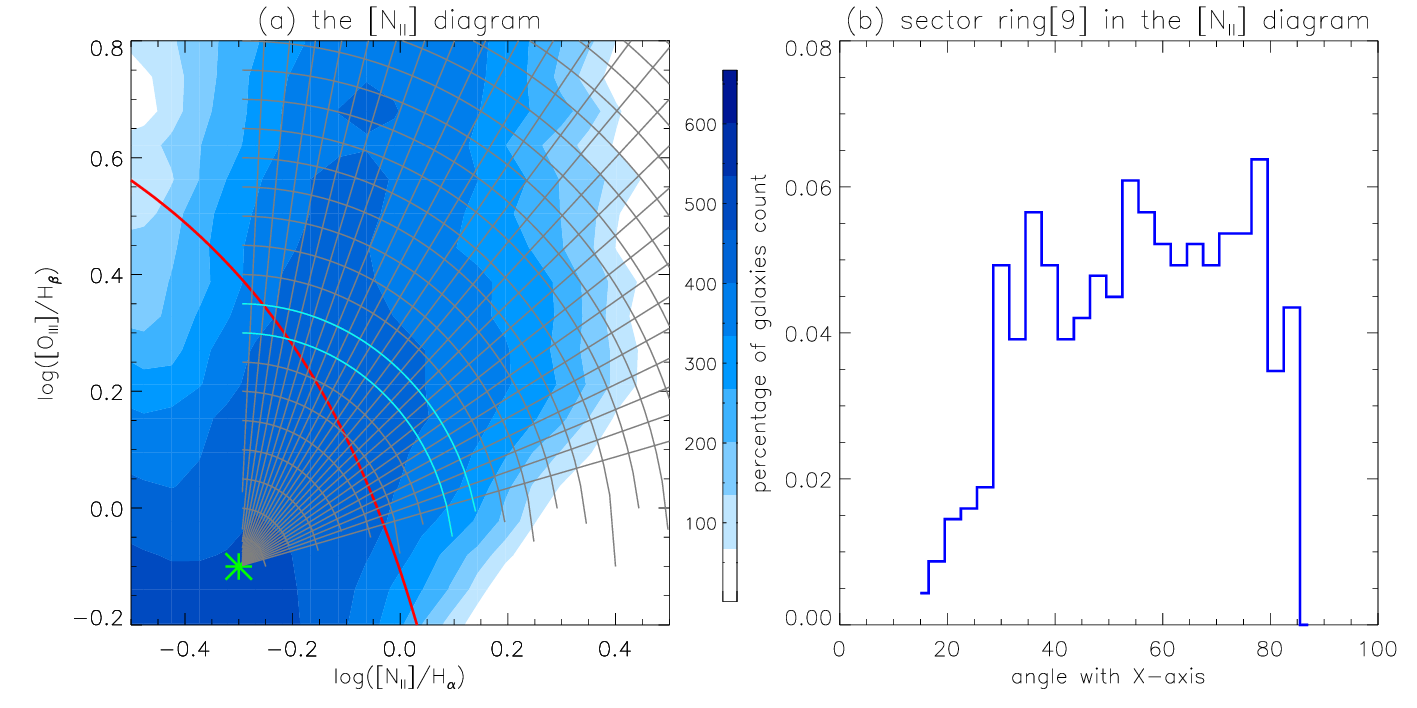}
	\caption{This figure demonstrates the complete process for defining the S-L line in the 
    [N~{\sc ii}] BPT diagrams using the method proposed in \citet{2006MNRAS.372..961K}. (a) shows
    the result of defining the S-L line in the [N~{\sc ii}] BPT diagram using the method proposed
    in \citet{2006MNRAS.372..961K}. The huge green asterisk represents the empirical base point. 
    The boundaries of sector ring[9] are marked in cyan. The solid red line is the Ke01 line. The 
    contour filled with bluish colors represents the results for the sample of our collected galaxies. 
    The corresponding number densities of the different colors are shown in the colorbar on the right
    side of the panel. (b) shows the dependence of the percentage of galaxies count within each 
    Ke-bin in sector ring[9] relative to the total count of galaxies in the entire sector ring on 
    the angle with the X-axis.   }        
	\label{fig:Fai-n}   
\end{figure*}

	When applying the above method in the [N~{\sc ii}] BPT diagram to determine the S-L line, 
the empirical base point coordinates are (log(\nii), log(\oiii)) = (-0.3, -0.1). The step between 
each sector ring is 0.05 dex, and the straight lines are also calibrated every 3 degrees starting 
from an angle of 15 degrees with the X-axis. However, there are no clear double-peaked features 
in the dependence of the percentage of galaxies count within each Ke-bin relative to the total 
count of galaxies in the corresponding entire sector ring on the angle with the X-axis, as 
illustrated by the example shown in Fig.~\ref{fig:Fai-n}. That is why a new independent method is 
proposed to determine an improved S-L line in the [N~{\sc ii}] BPT diagram.

\end{document}